\documentclass[a4paper,11pt]{article}
\usepackage{geometry}
\usepackage{a4wide,slashbox}
\usepackage{graphicx}
\usepackage{epsf}
\usepackage{cite}
\usepackage{multirow,tabularx}
\usepackage{appendix}
\usepackage{graphicx,amsmath,amsfonts,amssymb,amsthm,euscript,braket,xcolor}
\newcommand{\be}{\begin{equation}}
\newcommand{\ee}{\end{equation}}

\newcommand{\Rmnum}[1]{\expandafter\@slowromancap\romannumeral #1@}
\newcommand{\bea}{\begin{eqnarray}}
\newcommand{\eea}{\end{eqnarray}}

\newcommand{\p}{\partial}

\numberwithin{equation}{section}

\begin{document}

\title{\bf Confining gauge theories and holographic entanglement entropy with a magnetic field }
\author{\textbf{David Dudal$^{a,b}$}\thanks{david.dudal@kuleuven.be},
\textbf{Subhash Mahapatra$^{a}$}\thanks{subhash.mahapatra@kuleuven.be},
 \\\\\
\textit{{\small $^a$ KU Leuven Campus Kortrijk -- KULAK, Department of Physics, Etienne Sabbelaan 51 bus 7800,}}\\
\textit{{\small 8500 Kortrijk, Belgium}}\\
\textit{{\small $^b$           Ghent University, Department of Physics and Astronomy, Krijgslaan 281-S9, 9000 Gent, Belgium}}}
\date{}

\maketitle
\abstract{We consider the soft wall model for a heuristic holographical modelling of a confining gauge theory and discuss how the introduction of a (constant) magnetic field influences the (de)confinement phase structure. We use the entanglement entropy as a diagnostic tool in terms of the length of an entangling strip geometry. Due to the anisotropy introduced by the magnetic field, we find that the results depend on the orientation of the strip relative to the field. This allows to identify a richer, anisotropic, interplay between confinement and a magnetic field than possibly can be extracted from a more standard order parameter as, for example, the Polyakov loop expectation value.}

\section{Introduction}
It is by now well accepted that the gauge/gravity duality can provide important information into the physics of strongly coupled systems where a perturbative analysis may not be effective \cite{Maldacena,Klebanov9802,Witten9802}. The gauge/gravity duality in its approximate form maps a theory of classical gravity in AdS space to a strongly coupled conformal field theory, which lives at the boundary of the AdS space in a lower dimension. This mapping has been quite useful to understand and describe the physics of strongly coupled many body systems and there are indications that realistic predictions might be acquired from it. Two of the most important applications of the gauge/gravity duality that have received wide attention, and are a topic of this paper, are entanglement entropy and (the phase structure of) quantum chromodynamics (QCD).\\

The gauge/gravity duality provides many important concepts and associated tools to probe the properties of gauge theories at strong coupling. A key example of such concept is the entanglement entropy. A holographic proposal for the entanglement entropy has been conjectured in the milestone paper \cite{Ryu0603}\cite{Ryu0605}, with more recent proofs of the conjecture, generally applicable or not, in \cite{Casini:2011kv} or \cite{Lewkowycz:2013nqa}. The proposal geometrizes the notion of entanglement entropy and has been extensively tested for a variety of systems. It has been the subject of intense investigation during the last decade and has found applications in many diverse areas of physics. For example, holographic entanglement entropy has been used to study quantum quenches in strongly coupled systems \cite{Rangamani1512,Albash1008,Nozaki1302}, in black hole physics \cite{Solodukhin0606,Solodukhin1104}, to understand phase transitions in bulk and boundary theories \cite{Johnson1306,Caceres1507,Dey1512,Albash1202,Dey1409}, to study out of equilibrium dynamics such as thermalization in quark-gluon plasma \cite{Balasubramanian1012,Balasubramanian1103,Galante1205,Caceres1205,Dey1510,Liu1305} etc. Indeed, understanding the structure of (holographic) quantum entanglement has emerged as a fundamental question to address various novel emergent phenomena, ranging from many body complex quantum systems, to quantum phase transitions, to black holes. For reviews on this topic, let us refer to \cite{Nishioka0905,Rangamani1609,sachdev,Osborne,Latorre:2009zz}.\\

Another important application of the gauge/gravity duality is that it provides access to QCD at strong coupling \cite{Witten9803}. Normally, the boundary field theory in the context of gauge/gravity duality is subject to conformal invariance.  However, it is also well known that in order to describe realistic QCD from the gauge/gravity duality viewpoint, we need to break this conformal symmetry as the non-zero running of the coupling constant in real QCD generally breaks the conformal symmetry and via dimensional transmutation, a mass gap is formed. Several gravity models, rooted in string theory, that explicitly break the conformal symmetry have been constructed in the recent past and are commonly known as ``top down'' AdS/QCD models \cite{Polchinski0003,Sakai0412,Sakai0507,Kruczenski0311,Karch0205,Klebanov0007}.\\

In this work, we are interested in more phenomenological (``bottom-up'') models of AdS/QCD  where one constrains the gravity theory by hand as to reproduce the desirable features of the boundary theory resembling QCD, without actually deriving them from a consistent truncation of an underlying string theory \cite{Erlich,Paula,Gursoy, GursoyI,GursoyII,Herzog0608}. In particular, we are interested in the soft wall model description of AdS/QCD \cite{Karch0602,Karch1012}. In the soft wall models, one introduces an additional dilaton field that explicitly breaks the conformal symmetry in the IR regime. These models are quite useful in deriving many properties of real QCD, such as a reasonable meson mass spectrum, linear Regge trajectories etc. These models can also generate essential properties of chiral symmetry breaking and confinement, see e.g.~\cite{Erlich,Colangelo:2011sr,Dudal1511,Gherghetta:2009ac}.\\

However, they face a few drawbacks as well. For example, the soft wall models fail to reproduce the area law of the Wilson loop expectation value \cite{Karch1012}, they fail to produce a chiral condensate that is also non-zero for zero bare quark mass \cite{Colangelo:2011sr,Gherghetta:2009ac,Dudal1511} or the dilaton is added by hand, so the Einstein equations of motion are not explicitly satisfied. Although it is reasonable to say that connections to real QCD via the gauge/gravity duality appear to be limited in some aspects, it is even so important to explore this quest and further our understanding towards the main aim of connecting to genuine QCD. In fact, more involved wall models are on the market that remedy one or another of the aforementioned shortcomings \cite{Gursoy,GursoyI,GursoyII,Paula,Gherghetta:2009ac,Li:2016gfn,Lindgren:2015lia}.\\

The study of entanglement entropy in (confining) gauge theories has attracted a lot of attention lately, both from holographic as well as from non-holographic point of view. In holography, this question was first addressed in \cite{Klebanov0709}, which first generalised the entanglement entropy prescription of \cite{Ryu0603} to non-conformal field theories and then found that for gravity backgrounds, which are holographically dual to confining gauge theories, the entanglement entropy shows a first order phase transition upon varying the size of the entangling surface (a strip of length $\ell$). In particular, depending on the value of $\ell$, they found two minimal surfaces for the entanglement entropy: a connected and a disconnected one. The connected surface was found to have lower entanglement entropy below a certain critical $\ell_{crit}$ while the disconnected one was favoured above that value. Importantly, it was shown that the holographic entanglement entropy scales as $N^2$ for small $\ell$, and as $N^0$ for large $\ell$ . This resembles the expected characteristic features of (de)confining theories --- with $\ell$ playing the role of the inverse temperature. Indeed, below the critical temperature the coloured (gluon) degrees of freedom are confined, counting for $\mathcal{O}(N^0)$ degrees of freedom, while above that temperature, the deconfined degrees of freedom count as $\mathcal{O}(N^2)$. This analysis therefore provides a strong indication for the entanglement entropy as a non-local probe to diagnose confinement. Several authors then generalized this idea to a variety of confining systems and found similar results \cite{Kola1403,Kim,Lewkowycz,Ghodrati,Fujita0806}. It is also worthwhile mentioning that the bulk Ryu-Takayanagi prescription relates to the replica trick to compute the boundary entanglement entropy \cite{Lewkowycz:2013nqa}, whereof it was recently shown, at least for the Abelian case, that this entanglement entropy definition corresponds to the extended Hilbert space definition. Other (inequivalent) definitions exist, including one at the operational level which amends to a measurable notion of entanglement in gauge theories, which do not give the same results at the end, see \cite{Soni:2015yga,Soni:2016ogt} and also \cite{VanAcoleyen:2015ccp}. The replica trick/extended Hilbert space entanglement entropy is in itself a non-measurable quantity, nonetheless it was motivated in \cite{Soni:2016ogt} to be the correct definition after all. To end this paragraph, let us mention that the main features of the holographic entanglement entropy of a confining gauge theory have received numerical confirmation from the lattice papers \cite{Buividovich:2008gq,Buividovich:2008kq,Itou:2015cyu}.\\

Till now, to the best of our knowledge, all the works that have appeared in the literature and which are probing confinement using the holographic entanglement entropy have been carried out in the absence of background fields. However, it is by now well appreciated that in particular a magnetic field plays an important role in the QCD realm, given its creation during a non-central heavy ion collision and its ensuing appearance during the early stages of the quark-gluon plasma phase \cite{Kharzeev:2007jp,Skokov:2009qp,Bzdak:2011yy,Deng:2012pc,Tuchin:2013apa,Tuchin:2013ie,McLerran:2013hla}. We will model the magnetic field as $\vec{B}=B\vec{e}_z$ with $B$ constant. Also confinement physics is quite sensitive to the magnetic field, see in particular \cite{Johnson,Rougemont:2015oea,Callebaut1303,Bayona,Dudal1410,Dudal1510,Mamo1501,Dudal1511,McInnes:2015kec,Fang:2016cnt,Li:2016gtz,Li:2016gfn} for a few holographically oriented works, \cite{Bali:2011qj} for a breakthrough lattice QCD study and \cite{Kharzeev:2012ph,Miransky:2015ava} for recent reviews and many more references. For example, the critical temperature of the confinement/deconfinement phase transition was found to decrease under the influence of a magnetic field in the hard wall model \cite{Mamo1501}. Later in \cite{Dudal1511,Li:2016gfn}, the same result was shown to exist in the soft wall model. Similarly, inverse magnetic catalysis is an important characteristic property of the QCD chiral transition. Also, the magnetic field can strongly influence the meson spectrum \cite{Chernodub:2010qx,Chernodub:2011mc,Callebaut1105,Callebaut1309,Ali,Bu:2012mq}. For these reasons, it is important to investigate the possible effects that a background magnetic field might cause in the intertwinement of entanglement entropy and confinement. In this paper we exactly initiate such study.\\

However, in order to do so we need to take into account the backreaction of the magnetic field on the bulk geometry. Here we will rely on the results of \cite{Hoker0908,Hoker0911}, where a magnetised pure AdS background in the bulk was obtained after solving the Einstein-Maxwell system in the limit of small constant magnetic field with asymptotic AdS boundary conditions. Our strategy in this work is therefore to take the magnetized AdS background of \cite{Hoker0908,Hoker0911}, supplemented with a dilaton field put in by hand~\footnote{This is nothing else than the magnetic version of the working hypothesis of the original soft wall model \cite{Karch0602}, as proposed in \cite{Dudal1511}.} to model the soft wall confinement phase and then use the prescription of \cite{Klebanov0709} to study the entanglement entropy. This strategy leads to multiple tunable parameters in our model by construction which, as we will see later on, provides a far richer phase structure of the entanglement entropy compared to \cite{Klebanov0709}. These parameters are the magnetic field $B$, the dilaton scale factor $c$ and an additional length parameter $\ell_c$. The latter unavoidably enters in the magnetized AdS background and leads to several non-trivial effects on the entanglement entropy. In fact, the same parameter was already shown in \cite{Dudal1511} to play also a pivotal role in establishing the inverse magnetic catalysis in the deconfinement phase transition via the Hawking-Page analysis. It is interesting to remark here that the so-called ``improved holographic QCD models'' of \cite{Gursoy:2016ofp} recently saw an extension to the magnetic field case and also there, an extra parameter is necessarily introduced to play a key role in the phase diagram for both deconfinement and chiral transition. As the metric of \cite{Gursoy:2016ofp} is a purely numerical construct, we will not use it here to study the entanglement entropy, as this would technically be rather challenging.\\

Importantly, in the presence of  a magnetic field $\vec{B}$, there are several possibilities to choose the entangling surface. Clearly, we can thus expect to see an anisotropic footprint in the corresponding entanglement entropies. This is interesting, as it will allow to see also anisotropic features in the confinement properties of gauge theories. It is impossible to discover such properties in the Polyakov loop expectation value, a standard order parameter for the deconfinement transition which is insensitive to the direction of the (constant) magnetic field $\vec{B}$, it does depend however on its magnitude $B$. Though, interesting lattice evidence in favour of sizeable anisotropies in the string tension between two heavy quarks was provided for in \cite{Bonati:2014ksa,Bonati:2016kxj} by measuring expectation values of Wilson loops which, in contrast with a Polyakov loop, can be given a relative and variable orientation with respect to the magnetic field. The extracted string tension between heavy quarks becomes dependent on the magnetic field (orientation), both at zero and finite temperature, an observation that is of great phenomenological interest as the string tension is tightly correlated with the properties of (heavy) quark bound states such as charmonium, and as such the latter can be expected to be rather susceptible to the presence of a magnetic field, a fact supported by various sources \cite{Dudal1410,Machado:2013rta,Alford:2013jva,Cho:2014loa,Bonati:2015dka,Sadofyev:2015hxa,Fukushima:2015wck,Suzuki:2016kcs,Yoshida:2016xgm,Hattori:2016emy,Suzuki:2016fof}. As such, it is again clear that further understanding confinement in an anisotropic setting is important, both from theoretical as well as phenomenological viewpoint, see also \cite{Miransky:2002rp,Ozaki:2015yja}. For the record, let us mention that the effects of a magnetic field on the Wilson loop were also considered in \cite{Rougemont:2014efa}, albeit for the (naturally deconfined) $\mathcal{N}=4$ SYM case, while anisotropic entanglement entropy has been studied in a different context in \cite{Ecker:2015kna}. Some other general aspects of
anisotropic holographic quark-gluon plasmas have been studied in \cite{Giataganas:2012zy,Fadafan:2013bva}.  \\

For simplicity, we will choose the entangling surface in a direction either parallel or perpendicular to the magnetic field. This entails the possibility of two critical lengths, $\ell_{crit}^{\parallel}$ and $\ell_{crit}^{\perp}$, in our model at which the entanglement entropy shows a first order phase transition upon varying the size of the entangling surface. We find that this is indeed the case. A qualitative dependence of $\ell_{crit}^{\parallel}$ and $\ell_{crit}^{\perp}$ on $B$, $c$ and $\ell_c$ is shown in Table~\ref{tablelcrit}. Our main results are summarized as follows:
\begin{itemize}
 \item There is a maximum length $\ell_{max}$ above which the connected solution does not exist, only the disconnected surface remains. We find two such lengths, $\ell_{max}^{\parallel}$ and $\ell_{max}^{\perp}$, respectively for the parallel and perpendicular entangling strips. We find that both $\ell_{max}^{\parallel}$ and $\ell_{max}^{\perp}$ depend non-trivially on the model parameters $B$, $c$ and $\ell_c$. In particular, they can be a monotonic or non-monotonic function of the magnetic field and $\ell_{max}^{\parallel}$ can be larger or smaller than $\ell_{max}^{\perp}$, depending on the values of these parameters.

 \item With a background magnetic field, we again recover a first order phase transition in the entanglement entropy. We find that the connected surface displays a lower entanglement entropy below a certain critical length $\ell_{crit}$ while the disconnected configuration shows lower entanglement entropy above that value. Again, we find two such critical lengths, $\ell_{crit}^{\parallel}$ and $\ell_{crit}^{\perp}$. The behaviour of these critical lengths with respect to magnetic field depends again on the parameters $c$ and $\ell_c$. In particular, we find that $\ell_{crit}^{\parallel}$ and $\ell_{crit}^{\perp}$ can increase or decrease or even show non-monotonic behaviour with respect to the magnetic field. Further, we find that $\ell_{crit}^{\perp}$ can be larger or smaller than $\ell_{crit}^{\parallel}$. In particular, we find that $\ell_{crit}^{\perp}$ generally dominates $\ell_{crit}^{\parallel}$ for large $\ell_c$ whereas $\ell_{crit}^{\parallel}$ can be larger than $\ell_{crit}^{\perp}$ for very small values of $\ell_c$ .

 \item It is important to analyze similarities and differences between the critical temperature $T_{crit}\equiv \frac{1}{\ell_{crit}}$ and the critical temperature $T_{HP}$ of confinement/deconfinement phase transition obtained from the free energy analysis of the dual Hawking-Page transition \cite{Dudal1511,Herzog0608}. We can report that for a fixed $B$ and $c$, both $T_{crit}^\parallel$ and $T_{crit}^\parallel$ decrease for an increasing $\ell_c$. This behaviour is similar to the $T_{HP}$ result found in \cite{Dudal1511}, suggesting some kind of a close relationship between $T_{HP}$ and $T_{crit}$. However, there are also some notable differences. We find that the magnitudes of $c$ and $\ell_c$ at which $T_{HP}$ decreases with the magnetic field do not coincide very well with the corresponding values in $T_{crit}^\parallel$ (or $T_{crit}^\perp$).

\item We also establish the $\ell$ dependence of the (parallel and perpendicular) entropic $\cal C$-functions \cite{Itou:2015cyu}. Even with the background magnetic field, the $\cal C$-functions show the expected behaviour. In particular, we find that they decrease along the RG flow from UV to IR as we increase the length of the entangling strip and that they vanish at long distances. There is a sharp drop to zero when the critical lengths $\ell_{crit}^{\perp,\parallel}$ are approached, indicate of a phase transition.
\end{itemize}
The paper is organized as follows. In the next section, we will introduce our gravity model. In section 3, we will first highlight the entanglement entropy prescription of \cite{Klebanov0709} and briefly discuss their results to set the stage. We will then derive the necessary formulae for the entanglement entropy in the presence of  a background magnetic field. The numerical results of our computations for the entanglement entropy and entropic $\cal C$-function are presented and discussed in section 4. Finally, in section 5, we conclude this paper with an outlook to future research.

\section{Gravity setup}
In this section, we will briefly review the gravity dual of a confining gauge theory with a background magnetic field. We will follow the notation used in \cite{Dudal1511} and more details can thence be found there. We start from the Einstein-Maxwell action in the presence of a negative cosmological constant,
\begin{eqnarray}
&&S_{EM} = - \frac{1}{16 \pi G_5} \int \mathrm{d^5}x \sqrt{-g} \ \ \bigl[R + \frac{12}{L^2}
-\textit{F}_{MN}\textit{F}^{MN} \bigr]\,,
\label{action}
\end{eqnarray}
where $R$ is the Ricci scalar, $L$ is the AdS length related to the negative cosmological constant and $\textit{F}_{MN}$ is the electromagnetic field strength tensor. $G_5$ is the Newton constant in five dimensions. The equations of motion obtained by varying eq.~(\ref{action}) are
\begin{eqnarray}
&& R_{MN}-{1\over2}g_{MN}R-{6\over L^2} g_{MN} + \frac{1}{2} g_{MN}\textit{F}_{IJ}\textit{F}^{IJ}- 2\textit{F}_{MI}\textit{F}_{N}^{\ \ J} =0
\label{EinsteinEOM}
\end{eqnarray}
and
\begin{eqnarray}
&& \nabla_{M}\textit{F}^{MN}=0\,.
\label{maxwellEOM}
\end{eqnarray}
In order to solve eqs.~(\ref{EinsteinEOM}) and (\ref{maxwellEOM}) with constant magnetic field, one can choose the following ansatz for the metric
\begin{eqnarray}
ds^2=\frac{L^2}{r^2}\biggl(-f(r)dt^2 + \frac{dr^2}{f(r)} + h(r)(dx^2+dy^2) + q(r)dz^2 \biggr)\,,
\label{metric}
\end{eqnarray}
with $\vec{B}=B\vec{e}_z$, which breaks the rotation symmetry in the $z$-direction. One can easily check that a constant magnetic field configuration in the $z$-direction satisfies the Maxwell equations. Here, $r$ is the holographic radial coordinate and, in our notations, the asymptotic boundary is at $r=0$. Because of the complicated nature of the Einstein equations, which are non-linear coupled differential equations, analytic solutions are very difficult to obtain. However for small magnetic field, the Einstein equations can be solved perturbatively and the results were presented in \cite{Hoker0908,Hoker0911}. For magnetized AdS, the solution up to order $B^2$ can be written as~\footnote{In principle, the factor $L$ present in the $\ln (Lx)$ terms can be replaced by a random other length scale $\ell_Y$ as in \cite{Dudal1511}, but it drops out of physical quantities as discussed there. So, we have immediately set it equal to $L$ here.}
\begin{eqnarray}
&& f(r)=1+\frac{2}{3}\frac{B^2 r^4}{L^2}\ln{\bigl(\frac{r}{\ell_c} \bigr)} + \mathcal{O}(B^4)\,, \nonumber \\
&& q(r)=1+\frac{8}{3}\frac{B^2}{L^2} \int_{\infty}^{1/r} \mathrm{d}x \frac{\ln{(L \ x)}}{x^5} + \mathcal{O}(B^4)\,,  \nonumber \\
&& h(r)=1-\frac{4}{3}\frac{B^2}{L^2} \int_{\infty}^{1/r} \mathrm{d}x \frac{\ln{(L \ x)}}{x^5} + \mathcal{O}(B^4)\,.
\label{magnetizedAdS}
\end{eqnarray}
In the language of the AdS/CFT correspondence, the above magnetized thermal AdS solution is the gravity dual of the confined phase, up to a necessary modification. Indeed, to mimic QCD, following the soft wall philosophy of \cite{Karch0602}, the action density appearing in eq.~\eqref{action} is modified by hand with a dilaton prefactor, $e^{-\frac{c}{2}r^2}$ \footnote{It should be noted that the $c$ in our notation differs from the $c$ in \cite{Herzog0608,Karch0602,Dudal1511} by a factor of 2, this for later convenience.}, to smoothly cut off the AdS interior and to supplement the necessary scale breaking parameter $c$.\\

We will later use this magnetized soft wall AdS model to compute the holographic entanglement entropy of the confining gauge theory. However, before going into that, let us first briefly discuss a few silent features of eqs.~(\ref{EinsteinEOM})-(\ref{magnetizedAdS}). There is actually another important solution to eq.~(\ref{EinsteinEOM}), i.e.~the one with a horizon and thus corresponding to a black hole geometry. On the field theory side, this magnetized black hole solution corresponds to the deconfined phase. As shown in \cite{Dudal1511}, there can be a first order Hawking-Page phase transition from the magnetized AdS to the black hole metric as we increase the Hawking temperature. In particular, the free energy of the thermal AdS can become smaller/larger than the free energy of the black hole phase at low/high temperatures. This Hawking-Page phase transition on the field theory side corresponds to the famous confinement/deconfinement phase transition \cite{Witten9803}. Importantly, this transition was shown to exist for both soft as well as hard wall AdS/QCD models \cite{Herzog0608,BallonBayona:2007vp}, including in the magnetic field case \cite{Mamo1501,Dudal1511}.\\

In eq.~(\ref{magnetizedAdS}), we have introduced an additional length parameter~\footnote{This $\ell_c$ is not to be confused with the critical length $\ell_{crit}$ of the entangling strip surface.} $\ell_c$ in the geometry. For any choice of $\ell_c$, the above metric solves the Einstein equations up to order $B^2$ and therefore is a consistent solution of the Einstein equations. In \cite{Dudal1511}, the magnitude of the parameter $\ell_c$ was constrained by matching the $T=0$ chiral condensate estimate with that of the actual (lattice) QCD result which yields $\ell_c\approx 1~\text{GeV}^{-1}$. It was also shown that for $\ell_c  \gtrapprox 1~\text{GeV}^{-1}$, the critical temperature $T_{HP}$ of the confinement/deconfinement phase transition decreases with the magnetic field, which qualitatively agrees with the actual QCD behaviour \cite{Bali:2011qj}. However for $\ell_c$ much smaller than $1~\text{GeV}^{-1}$, the critical temperature was found to increase with the magnetic field. Clearly, $\ell_c$ is of physical relevance. It would therefore be interesting to see whether similar kind of results can be captured by the entanglement entropy. In particular, it would be instructive to compare the results of $T_{HP}$ with its analogue $T_{crit}$, which appears in the entanglement entropy analysis.

\section{The holographic entanglement entropy in confining backgrounds}
We now proceed to study the entanglement entropy of the boundary confining gauge theory in the presence of a background magnetic field. To compute the entanglement entropy we will use the prescription of \cite{Klebanov0709}, which is the generalization of Ryu-Takayanagi conjecture to non-conformal field theories. In this prescription, the EE between a spatial region $A$ and its complement is obtained by extremizing the following expression \footnote{A similar expression for the entanglement entropy previously also appeared in \cite{Ryu0605}.}
\begin{eqnarray}
S=\frac{1}{4G_{10}} \int_{\gamma}\mathrm{d}^5Y \mathrm{d}^3\sigma \ e^{-\phi} \sqrt{\gamma_{ind}}\,,
\label{SEE}
\end{eqnarray}
where $G_{10}$ is the ten-dimensional Newton constant and $\gamma_{ind}$ is the induced metric on the bulk surface $\gamma$, which propagates from the asymptotic boundary to the bulk and shares the same boundary $\partial A$ of the subsystem $A$. In our notation, $\frac{\phi}{2}$ is the dilaton field and, as mentioned before, we will make the standard choice $\phi=c r^2$ for it throughout this paper. The value of $c$ can be fixed by matching the soft wall prediction for the lightest vector meson on its experimental value, $m_\rho=0.776~$GeV, leading to a value of $c\simeq 0.3~\text{GeV}^2$. $\vec{\sigma}$ coordinatizes the three-dimensional induced metric and $\vec{Y}$ coordinatizes the five-dimensional internal space (apart from the five-dimensional AdS space of eq.~(\ref{metric})). The internal space will not play any significant role here and will be suppressed from the text from here on~\footnote{One can think of eq.~(\ref{SEE}) as the entanglement entropy measured per unit of volume of the internal space.}.  The entanglement entropy is then computed by minimizing the above action over all surfaces that approach to $\partial A$ at the asymptotic boundary.\\

As was considered in \cite{Klebanov0709}, here too, we consider the subsystem $A$ as a straight strip of length $\ell$. However with a background magnetic field, there are  separate possibilities to select the subsystem $A$. We will either choose $A$ parallel to or perpendicular to the applied magnetic field. As we will see below, for both these configurations the equations for entanglement entropy and length $\ell$ are different and will thus lead to different results.\\

However, before going to discuss each case separately, let us first briefly survey some silent features of \cite{Klebanov0709} for which parallel and perpendicular surfaces coincide. In \cite{Klebanov0709}, it was shown that for a given $\ell$ there are two local minimal surfaces emerging from eq.~(\ref{SEE}): a disconnected and a connected one. The disconnected surface consists of two lines which are separated by distance $\ell$ while having lower entanglement entropy for larger values of $\ell$. On the other hand, the connected surface resembles more of a tube connecting the two endpoints of the strip and it has lower entanglement entropy for smaller values of $\ell$. Importantly, it was found that the connected surface has no solution above a maximum length $\ell_{max}$ and that there is a phase transition from connected to disconnected surfaces as we steadily increase the length $\ell$. The phase transition was shown to occur at a critical length $\ell_{crit}<\ell_{max}$, below (above) which connected (disconnected) surface have a smaller entanglement entropy than the disconnected (connected) surface. This phase transition between the two surfaces was interpreted as characteristic of confining gauge theories, related to the aforementioned counting of relevant degrees of freedom at large $N$.\\

As mentioned before, due to the magnetic field, there are several additional relevant parameters in the theory (next to $B$ itself, also $c$ and $\ell_c$), and we will investigate how these additional parameters influence the structure of the above mentioned phase transition.  The boundary theory will now possess two critical lengths: one for the parallel  and one for the perpendicular surface and they can non-trivially depend on these parameters.

\subsection{Parallel entangling surface}
In this case, we choose our subsystem $A$ in the $z$-direction and the domain $-\ell/2\leq z \leq \ell/2$, $0\leq x \leq L_x$ and $0\leq y \leq L_y$ to define the strip geometry on the boundary. In order to find the minimal area solution for the connected surface we parameterize the surface $\gamma$ by
$z=z(r)$, with inverse $r=r(z)$. Now substituting eq.~(\ref{metric}) into eq.~(\ref{SEE}), we get
\begin{eqnarray}
S^{\parallel}_{conn}=\frac{L_x L_y L^3}{4G_{10}} \int_{-\ell/2}^{\ell/2}\mathrm{d}z \ \frac{e^{-\phi(r)}h(r)}{r^3}\sqrt{q(r)+\frac{r'^2}{f(r)}}\,.
\end{eqnarray}
As the corresponding Lagrangian $\mathcal{L}$ does not directly depend on $z$, the ``Hamiltonian'' $\mathcal{H}$ is conserved,
$\frac{\p}{\p z}\mathcal{H}=\frac{\p}{\p z}[r'\frac{\delta \mathcal{L}}{\delta r'}-\mathcal{L}]=0$. This leads to the following equation,
\begin{eqnarray}
e^{\phi(r)} \sqrt{q(r)+\frac{r'^2}{f(r)}}=e^{\phi(r_*)} \frac{r_{*}^3}{r^3} \frac{q(r)h(r)}{h(r_*)\sqrt{q(r_*)}}\,,
\label{minimalzpara}
\end{eqnarray}
where $r_*$ is the turning point of the minimal area surface at which $r'(z)|_{r=r_*}=0$. Finally, substituting eq.~(\ref{minimalzpara}) into eq.~(\ref{SEE}), we obtained the expression for the entanglement entropy for the connected surface as
\begin{eqnarray}
&& S^{\parallel}_{conn}=\frac{2L_x L_y L^3}{4G_{10}} \int_{\varepsilon}^{r_*}\mathrm{d}r \ \frac{r_{*}^{3}}{r^3}\frac{e^{-2\phi(r)}q(r)h^{2}(r)}{\sqrt{\bigl[r_{*}^{6}e^{-2\phi(r)}q(r)h^2(r)-r^{6}e^{-2\phi(r_*)}q(r_*)h^2(r_*) \bigr]f(r)q(r)}} \nonumber \\
&& \hspace{1cm} = \frac{2L_x L_y L^3}{4G_{10}} (\mathcal{S}^{\parallel}_{conn}+\frac{1}{2\varepsilon^2})\,,
\label{SEEzparacon}
\end{eqnarray}
where $r=\varepsilon$ defines the short distance $UV$ cutoff. It is introduced to regularize the entanglement entropy, which is diverging as ``too many'' UV degrees of freedom are living near to the strip boundary. Since the finite part $\mathcal{S}^{\perp}_{conn}$ of the entanglement entropy is independent of the cutoff, this is the quantity which is physically most relevant. However, in most part of this paper we will deal with the difference in entanglement entropy anyhow where the diverging parts trivially cancel out.\\

Correspondingly, the length of the strip surface for the connected solution as function of $r_*$ is
\begin{eqnarray}
\ell^{\parallel}_{conn} = 2 \int_{\varepsilon}^{r_*}\mathrm{d}r \ \frac{e^{-\phi(r_*)}r^{3}\sqrt{q(r_*)h^{2}(r_*)}}{\sqrt{\bigl[r_{*}^{6}e^{-2\phi(r)}q(r)h^2(r)-r^{6}e^{-2\phi(r_*)}q(r_*)h^2(r_*) \bigr]f(r)q(r)}}\,,
\label{lengthzparacon}
\end{eqnarray}
which is a finite quantity. Similarly, the entanglement entropy for the disconnected surface is given by~\footnote{This can be obtained by formally taking the $r_*\to\infty$ limit of the first line of eq.~\eqref{SEEzparacon}.}
\begin{eqnarray}
&& S^{\parallel}_{disc}=\frac{2L_x L_y L^3}{4G_{10}} \int_{\varepsilon}^{\infty}\mathrm{d}r \ \frac{e^{-\phi(r)}}{r^3}\sqrt{\frac{h^2(r)}{f(r)}} \nonumber \\
&& \hspace{1cm} = \frac{2L_x L_y L^3}{4G_{10}} (\mathcal{S}^{\parallel}_{disc}+\frac{1}{2\varepsilon^2})\,,
\label{SEEzparadis}
\end{eqnarray}
which is independent of $r_*$ and therefore of $\ell^{\parallel}_{conn}$ as well.

\subsection{Perpendicular entangling surface}
In this case, we can choose our subsystem $A$ in the $x$-direction and the domain $-\ell/2\leq x \leq \ell/2$, $0\leq y \leq L_y$ and $0\leq z \leq L_z$ to define the strip geometry on the boundary. In this case, we parameterize the connected surface $\gamma$ by $x=x(r)$ and $r=r(x)$ as inverse. Now putting eq.~(\ref{metric}) into eq.~(\ref{SEE}), we get
\begin{eqnarray}
S^{\perp}_{conn}=\frac{L_y L_z L^3}{4G_{10}} \int_{-\ell/2}^{\ell/2}\mathrm{d}x \ \frac{e^{-\phi(r)}}{r^3}\sqrt{q(r)h(r)\big[h(r)+\frac{r'^2}{f(r)}\bigl]}\,.
\end{eqnarray}
Using the same method as for the parallel case, we get the following equation,
\begin{eqnarray}
e^{\phi(r)} \sqrt{q(r)h(r)\big[h(r)+\frac{r'^2}{f(r)}\bigl]}=e^{\phi(r_*)} \frac{r_{*}^3}{r^3} \frac{q(r)h^2(r)}{\sqrt{q(r_*)h^2(r_*)}}\,,
\label{minimalxpara}
\end{eqnarray}
where $r_*$ is the turning point of the minimal area surface at which $r'(x)|_{r=r_*}=0$. Finally, substituting eq.~(\ref{minimalxpara}) into eq.~(\ref{SEE}), we obtain the expression for the entanglement entropy for the connected surface as
\begin{eqnarray}
&& S^{\perp}_{conn}=\frac{2L_y L_z L^3}{4G_{10}} \int_{\varepsilon}^{r_*}\mathrm{d}r \ \frac{r_{*}^{3}}{r^3}\frac{e^{-2\phi(r)}q(r)h^{2}(r)}{\sqrt{\bigl[r_{*}^{6}e^{-2\phi(r)}q(r)h^2(r)-r^{6}e^{-2\phi(r_*)}q(r_*)h^2(r_*) \bigr]f(r)h(r)}} \nonumber \\
&& \hspace{1cm} = \frac{2L_y L_z L^3}{4G_{10}} (\mathcal{S}^{\perp}_{conn}+\frac{1}{2\varepsilon^2})\,.
\label{SEExparacon}
\end{eqnarray}
The length of the strip for the connected solution as function of $r_*$ is
\begin{eqnarray}
\ell^{\perp}_{conn} = 2 \int_{\varepsilon}^{r_*}\mathrm{d}r \ \frac{e^{-\phi(r_*)}r^{3}\sqrt{q(r_*)h^{2}(r_*)}}{\sqrt{\bigl[r_{*}^{6}e^{-2\phi(r)}q(r)h^2(r)-r^{6}e^{-2\phi(r_*)}q(r_*)h^2(r_*) \bigr]f(r)h(r)}}\,.
\label{lengthxparacon}
\end{eqnarray}
Similarly, the entanglement entropy for the disconnected surface is given by
\begin{eqnarray}
&& S^{\perp}_{disc}=\frac{2L_y L_z L^3}{4G_{10}} \int_{\varepsilon}^{\infty}\mathrm{d}r \ \frac{e^{-\phi(r)}}{r^3}\sqrt{\frac{q(r)h(r)}{f(r)}} \nonumber \\
&& \hspace{1cm} = \frac{2L_y L_z L^3}{4G_{10}} (\mathcal{S}^{\perp}_{disc}+\frac{1}{2\varepsilon^2})\,,
\label{SEExparadis}
\end{eqnarray}
which is again independent of $r_*$ and $\ell^{\perp}_{conn}$.

\section{Results}
\begin{figure}[t!]
\begin{minipage}[b]{0.5\linewidth}
\centering
\includegraphics[width=2.8in,height=2.3in]{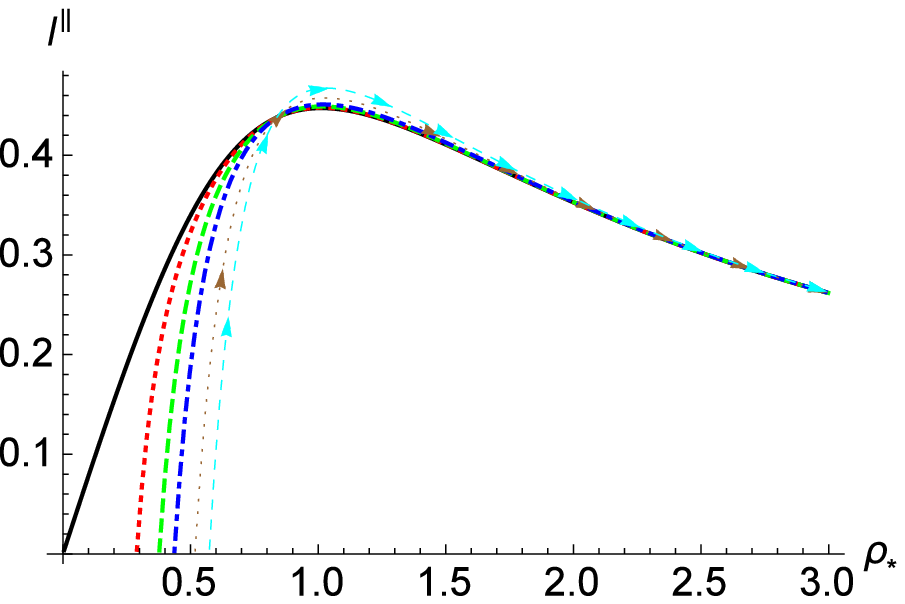}
\caption{ \small $\ell^\parallel$ as a function of $\rho_*$ for various $B$, with $\ell_c=1$. Here $c=2$ and (solid, black), (dot, red), (dash, green), (dot-dash, blue), (arrow-dot, brown) and (arrow-dash, cyan) curves correspond to $B=0$, $0.1$, $0.2$, $0.3$, $0.5$ and $0.7$ respectively. In units GeV.}
\label{lvsBc2lc1Zpara}
\end{minipage}
\hspace{0.4cm}
\begin{minipage}[b]{0.5\linewidth}
\centering
\includegraphics[width=2.8in,height=2.3in]{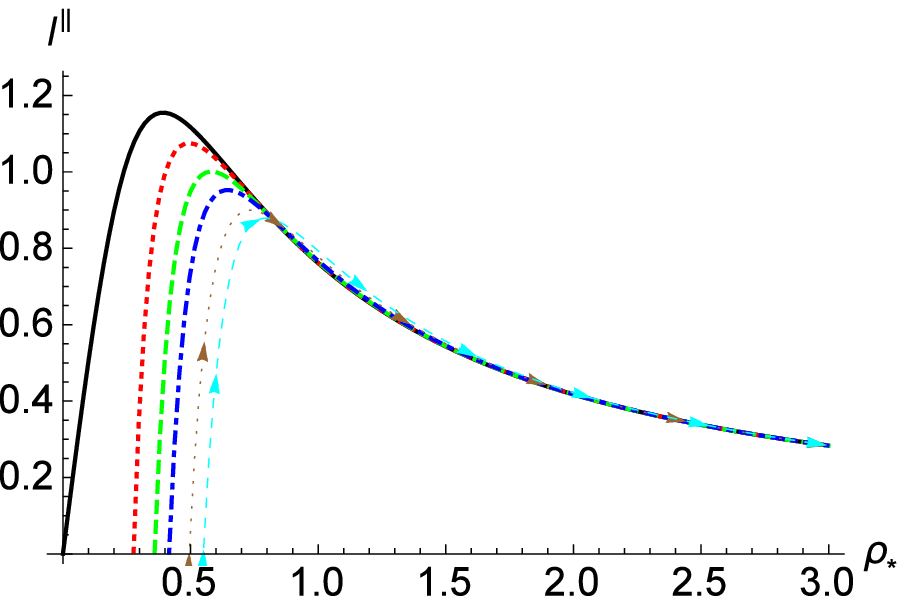}
\caption{\small $\ell^\parallel$ as a function of $\rho_*$ for various $B$, with $\ell_c=1$. Here  $c=0.3$ and (solid, black), (dot, red), (dash, green), (dot-dash, blue), (arrow-dot, brown) and (arrow-dash, cyan) curves correspond to $B=0$, $0.1$, $0.2$, $0.3$, $0.5$ and $0.7$ respectively. In units GeV.}
\label{lvsBcPt3lc1Zpara}
\end{minipage}
\end{figure}

In this section, we will present our results for the entanglement entropy. Since analytic results are hard to obtain for the connected surfaces, therefore, we will mainly focus on the numerical results. In order to solve eqs.~(\ref{SEEzparacon}), (\ref{lengthzparacon}), (\ref{SEExparacon}) and (\ref{lengthxparacon}) numerically, it turns out to be more convenient to use the coordinate $\rho=\frac{1}{r}$ to ensure stable numerics. This being said, closed analytic expressions can be found for the disconnected surfaces but these are cumbersome and not very illuminating after all.

\subsection{Parallel entangling surface}
Let us first analyze the entanglement entropy for a parallel entangling surface. The length $\ell^\parallel$ of the connected surface as function of $\rho_*$ for various values of the magnetic field $B$ and in decreasing order of $c$ is plotted in Figures \ref{lvsBc2lc1Zpara}-\ref{lvsBcPt3lc1Zpara}. Here we have fixed $\ell_c=1~\text{GeV}^{-1}$. In each Figure, (solid, black), (dot, red), (dash, green), (dot-dash, blue), (arrow-dot, brown) and (arrow-dash, cyan) curves correspond to $B=0$, $0.1$, $0.2$, $0.3$, $0.5$ and $0.7$ respectively. We observe that there are two solutions for a given $\ell$: one for small $\rho_*$ and one for large $\rho_*$. As we will see later, the one with larger $\rho_*$ corresponds to an actual local minimum of the entanglement entropy whereas the one with smaller $\rho_*$ corresponds to a saddle point.\\

\begin{figure}[t!]
\begin{minipage}[b]{0.5\linewidth}
\centering
\includegraphics[width=2.8in,height=2.3in]{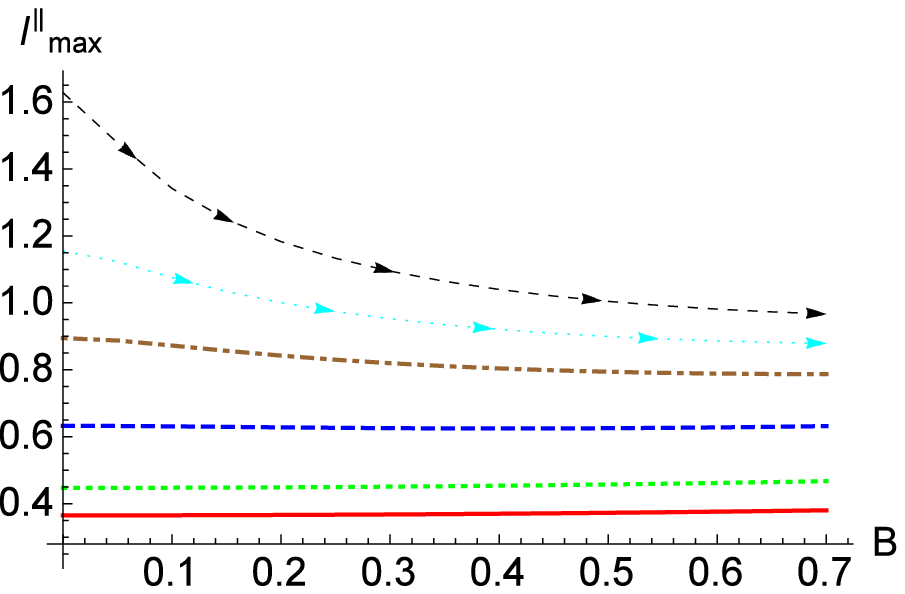}
\caption{ \small $\ell_{max}^{\parallel}$ as a function of $B$ for various $c$. Here $\ell_c=1$ and (solid, red), (dot, green), (dash, blue), (dot-dash, brown), (arrow-dot, cyan) and (arrow-dash, black) curves correspond to
$c=3$, $2$, $1$, $0.5$, $0.3$ and $0.15$ respectively. In units GeV.}
\label{lmaxvsBlc1Zpara}
\end{minipage}
\hspace{0.4cm}
\begin{minipage}[b]{0.5\linewidth}
\centering
\includegraphics[width=2.8in,height=2.3in]{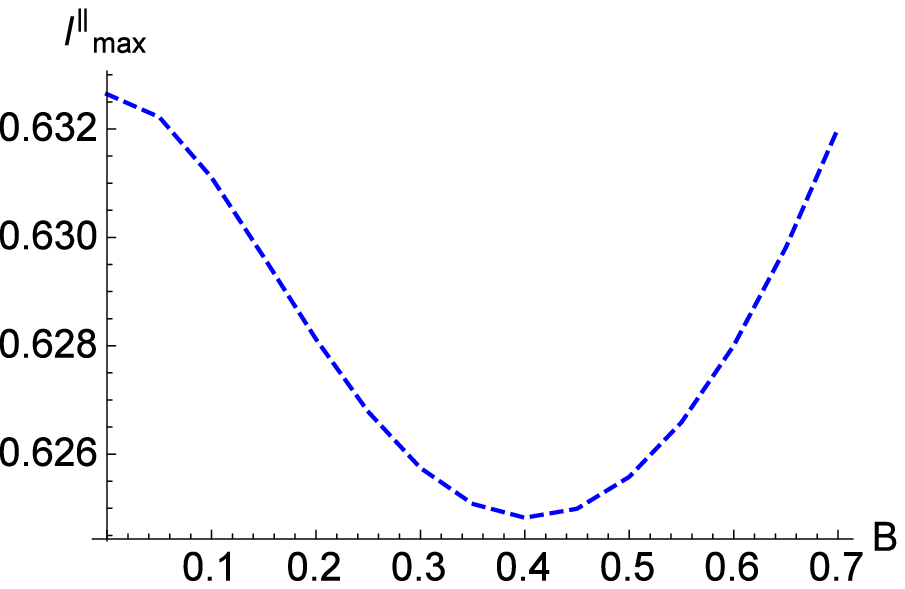}
\caption{\small $\ell_{max}^{\parallel}$ as a function of $B$ for $c=1$. In units GeV.}
\label{lmaxvsBc1lc1Zpara}
\end{minipage}
\end{figure}

As in the case of $B=0$, we see the occurrence of an $\ell_{max}^{\parallel}$ above which the connected solution does not exist. However for $B\neq0$, we now see that $\ell_{max}^{\parallel}$ depends quite non-trivially on both $B$ and $c$. For example, for $c=2~\text{GeV}^2$ (Figure \ref{lvsBc2lc1Zpara}), $\ell_{max}^{\parallel}$ increases with $B$ but as we decrease the value of $c$ the pattern reverses and $\ell_{max}^{\parallel}$ starts decreasing with $B$. This behaviour can be clearly seen in Figure \ref{lvsBcPt3lc1Zpara}, where we have used $c=0.3~\text{GeV}^2$. This suggests that there might be an intermediate value of $c$, for which $\ell_{max}^{\parallel}$ is not monotonic as a function of $B$. This is indeed the case as can be observed from Figures \ref{lmaxvsBlc1Zpara} and \ref{lmaxvsBc1lc1Zpara}, where the complete picture of the dependence of $\ell_{max}^{\parallel}$ on $B$ and $c$ is shown. We see that for $c=1~\text{GeV}^2$, $\ell_{max}^{\parallel}$ first decreases and then increases with magnetic field, indicating the non-monotonic behaviour of $\ell_{max}^{\parallel}$.
\begin{figure}[t!]
\begin{minipage}[b]{0.5\linewidth}
\centering
\includegraphics[width=2.8in,height=2.3in]{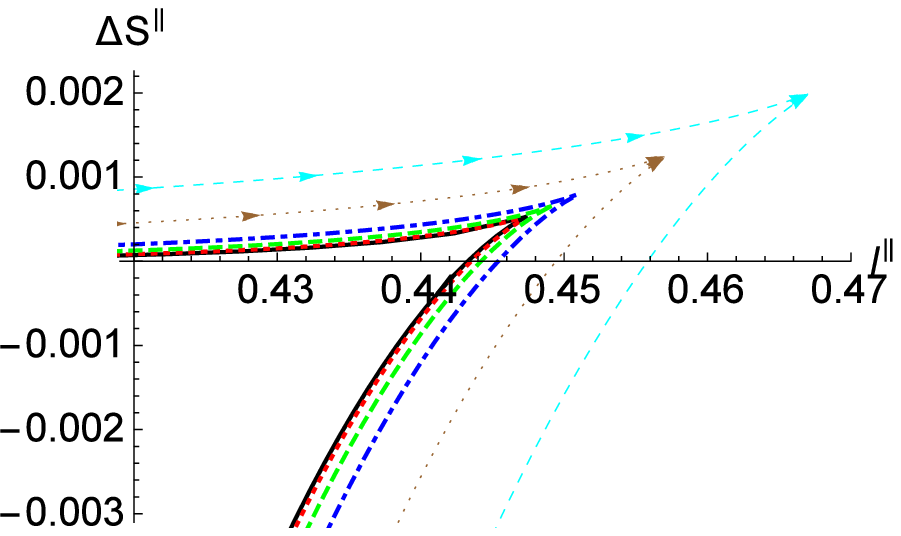}
\caption{ \small $\Delta S^{\parallel}$ as a function of $\ell^{\parallel}$ for various $B$, with $\ell_c=1$. Here $c=2$ and (solid, black), (dot, red), (dash, green), (dot-dash, blue), (arrow-dot, brown) and (arrow-dash, cyan) curves correspond to $B=0$, $0.1$, $0.2$, $0.3$, $0.5$ and $0.7$ respectively. In units GeV.}
\label{SvslvsBlc1c2Zpara}
\end{minipage}
\hspace{0.4cm}
\begin{minipage}[b]{0.5\linewidth}
\centering
\includegraphics[width=2.8in,height=2.3in]{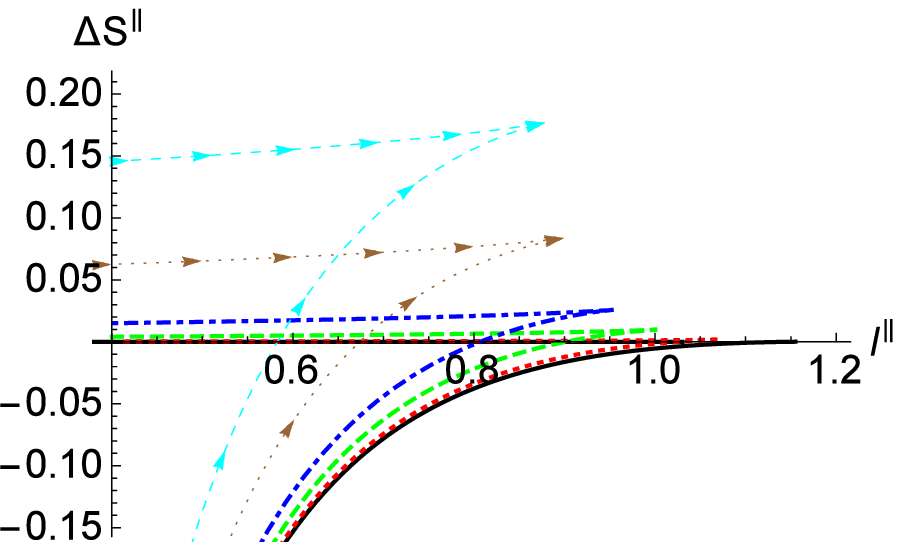}
\caption{\small $\Delta S^{\parallel}$ as a function of $\ell^{\parallel}$ for various $B$, with $\ell_c=1$. Here $c=0.3$ and (solid, black), (dot, red), (dash, green), (dot-dash, blue), (arrow-dot, brown) and (arrow-dash, cyan) curves correspond to $B=0$, $0.1$, $0.2$, $0.3$, $0.5$ and $0.7$ respectively. In units GeV.}
\label{SvslvsBlc1cPt3Zpara}
\end{minipage}
\end{figure}

Now, we move on to discuss the entanglement entropy itself. In order to do so, it is more convenient to consider the difference between connected and disconnected entanglement entropies~\footnote{In the numerical results presented here the constant prefactor $2 L_{x}L_{y}/4 G_{10}$ in the entanglement entropy expression has been suppressed.},
\begin{eqnarray}
\Delta S^{\parallel}=S^{\parallel}_{conn}-S^{\parallel}_{disc}\,.
\label{delSEEzpara}
\end{eqnarray}
$\Delta S^{\parallel}$ as function of $\ell^{\parallel}$ for various values of $B$ is shown in Figures \ref{SvslvsBlc1c2Zpara} and \ref{SvslvsBlc1cPt3Zpara}. In both these Figures, upper and lower lines correspond to smaller and larger branches of $\ell^{\parallel}$ respectively (see Figure \ref{lvsBc2lc1Zpara}). We see that the latter branch always has a lower entanglement entropy than the former one, indicating that it is a true minimum of the connected solution. In both Figures $\Delta S^{\parallel}$ can either be negative or positive depending on the value of $\ell^{\parallel}$. The former case occur for small values of $\ell^{\parallel}$, implying that the connected surface is the relevant solution of eq.~(\ref{SEE}) whereas the latter case occurs for large values of $\ell^{\parallel}$, implying that the disconnected surface becomes relevant. This corresponds to a phase transition from the connected to disconnected solution as we increase the strip length. The length at which this phase transition occurs defines the critical length $\ell_{crit}^{\parallel}$. One can see that $\ell_{crit}^{\parallel}$ is always less that $\ell_{max}^{\parallel}$, suggesting that this phase transition always occurs and it is of first order. This phase transition between the two solutions was suggested as characteristic for confining theories in \cite{Klebanov0709}. Here, we thus present evidence that a similar, albeit more intricate, phase transition structure exists in the presence of a background magnetic field as well.
\begin{figure}[t!]
\centering
\includegraphics[width=2.8in,height=2.3in]{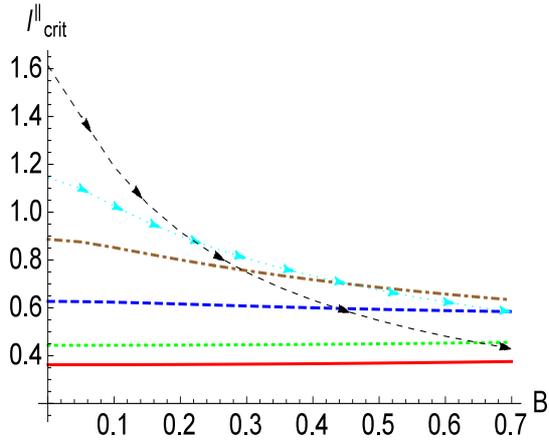}
\caption{ \small $\ell_{crit}^{\parallel}$ as a function of $B$ for various $c$. Here  $\ell_c=1$ and (solid, red), (dot, green), (dash, blue), (dot-dash, brown), (arrow-dot, cyan) and (arrow-dash, black) curves correspond to
$c=3$, $2$, $1$, $0.5$, $0.3$ and $0.15$ respectively. In units GeV.}
\label{lcritvsBlc1Zpara}
\end{figure}
\\
From Figures \ref{SvslvsBlc1c2Zpara} and \ref{SvslvsBlc1cPt3Zpara}, we see that $\ell_{crit}^{\parallel}$ depends quite nontrivially on $c$. For higher values of $c$, say $c=2~\text{GeV}^2$, $\ell_{crit}^{\parallel}$ increases with the magnetic field. However for lower values of $c$, it decreases with the magnetic field. A qualitative picture of the dependence of $\ell_{crit}^{\parallel}$ on $B$ and $c$ is shown in Figure \ref{lcritvsBlc1Zpara}. We find that although, for a fixed $c$, $\ell_{crit}^{\parallel}$ is a monotonic function of $B$, it shows non-monotonic behaviour as we vary $c$.

\begin{figure}[t!]
\begin{minipage}[b]{0.5\linewidth}
\centering
\includegraphics[width=2.8in,height=2.3in]{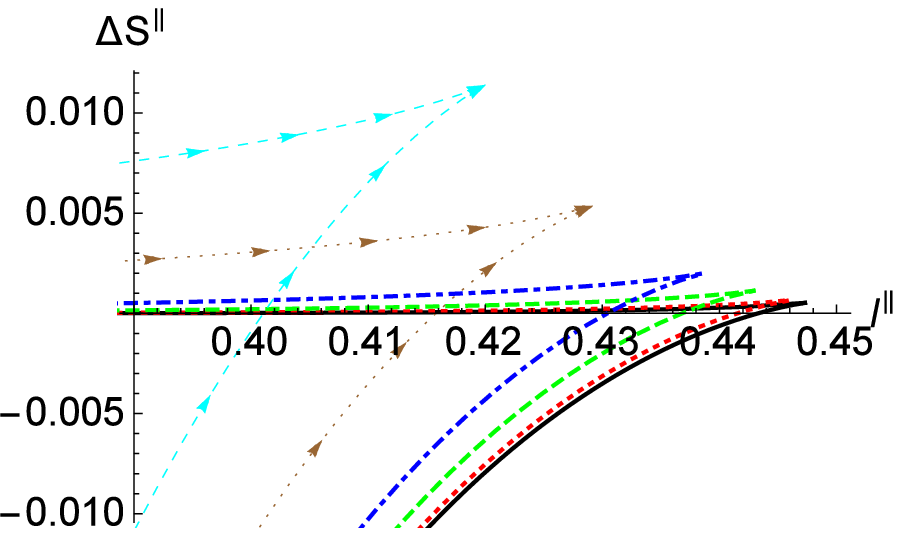}
\caption{ \small $\Delta S^{\parallel}$ as a function of $\ell^{\parallel}$ for various $B$, with $\ell_c=0.2$. Here $c=2$ and (solid, black), (dot, red), (dash, green), (dot-dash, blue), (arrow-dot, brown) and (arrow-dash, cyan) curves correspond to $B=0$, $0.1$, $0.2$, $0.3$, $0.5$ and $0.7$ respectively. In units GeV.}
\label{SvslvsBlcPt2c2Zpara}
\end{minipage}
\hspace{0.4cm}
\begin{minipage}[b]{0.5\linewidth}
\centering
\includegraphics[width=2.8in,height=2.3in]{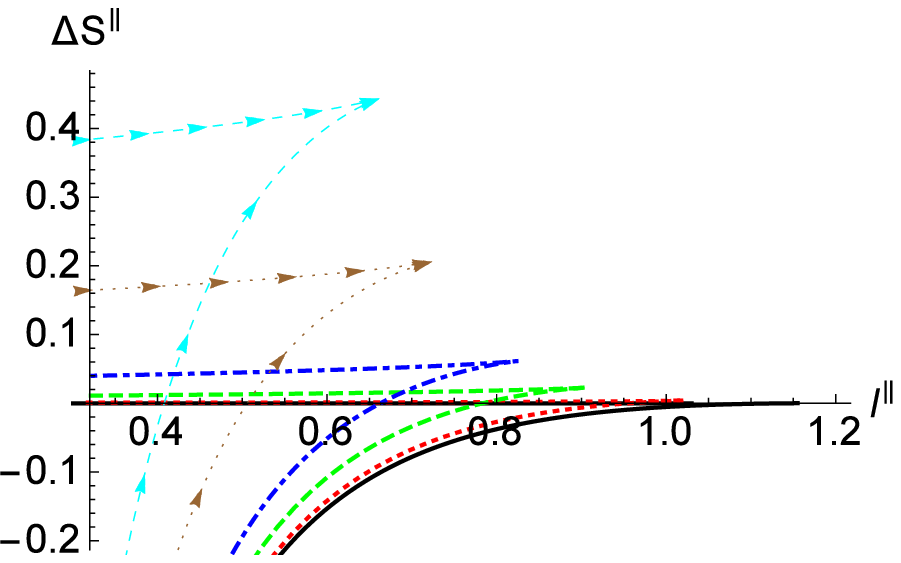}
\caption{\small $\Delta S^{\parallel}$ as a function of $\ell^{\parallel}$ for various $B$, with $\ell_c=0.2$. Here $c=0.3$ and (solid, black), (dot, red), (dash, green), (dot-dash, blue), (arrow-dot, brown) and (arrow-dash, cyan) curves correspond to $B=0$, $0.1$, $0.2$, $0.3$, $0.5$ and $0.7$ respectively. In units GeV.}
\label{SvslvsBlcPt2cPt3Zpara}
\end{minipage}
\end{figure}
\begin{figure}[t!]
\begin{minipage}[b]{0.5\linewidth}
\centering
\includegraphics[width=2.8in,height=2.3in]{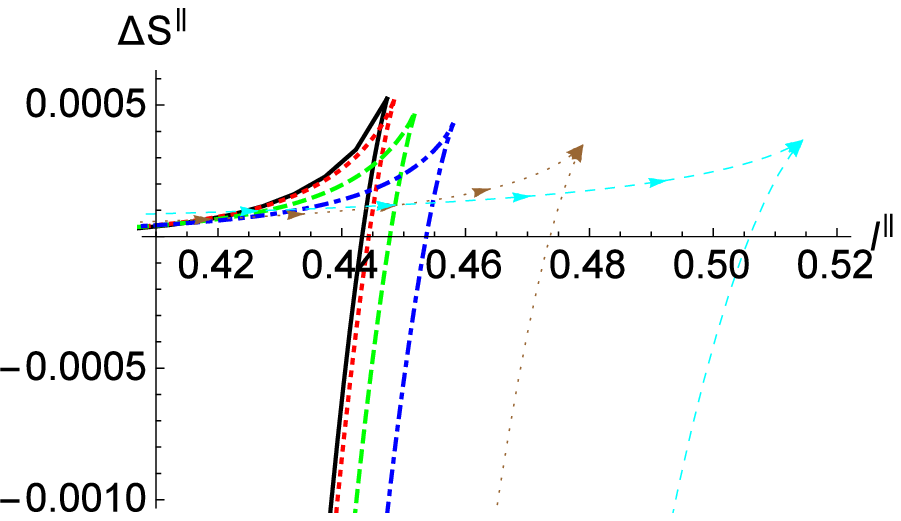}
\caption{ \small $\Delta S^{\parallel}$ as a function of $\ell^{\parallel}$ for various $B$, with $\ell_c=2$. Here $c=2$ and (solid, black), (dot, red), (dash, green), (dot-dash, blue), (arrow-dot, brown) and (arrow-dash, cyan) curves correspond to $B=0$, $0.1$, $0.2$, $0.3$, $0.5$ and $0.7$ respectively. In units GeV.}
\label{SvslvsBlc2c2Zpara}
\end{minipage}
\hspace{0.4cm}
\begin{minipage}[b]{0.5\linewidth}
\centering
\includegraphics[width=2.8in,height=2.3in]{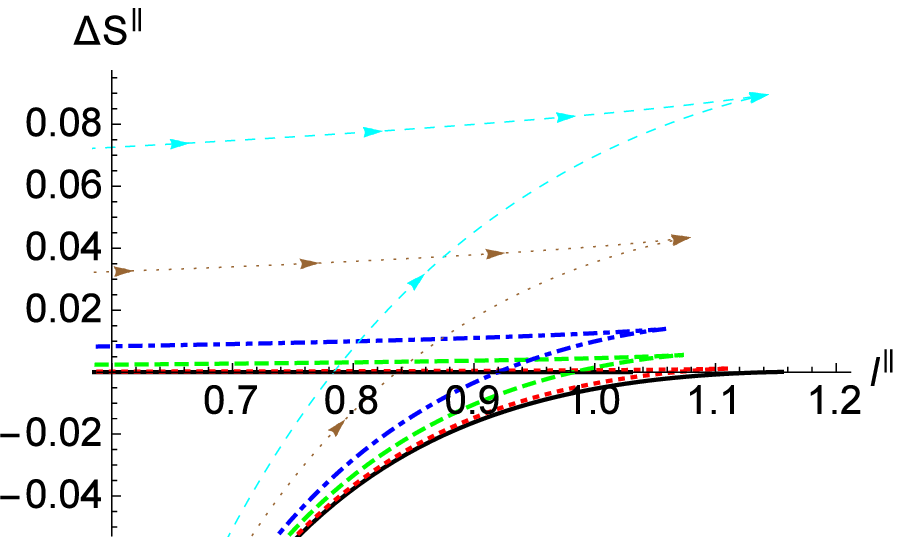}
\caption{\small $\Delta S^{\parallel}$ as a function of $\ell^{\parallel}$ for various $B$, with $\ell_c=2$. Here $c=0.3$ and (solid, black), (dot, red), (dash, green), (dot-dash, blue), (arrow-dot, brown) and (arrow-dash, cyan) curves correspond to $B=0$, $0.1$, $0.2$, $0.3$, $0.5$ and $0.7$ respectively. In units GeV.}
\label{SvslvsBlc2cPt3Zpara}
\end{minipage}
\end{figure}
\begin{figure}[t!]
\begin{minipage}[b]{0.5\linewidth}
\centering
\includegraphics[width=2.8in,height=2.3in]{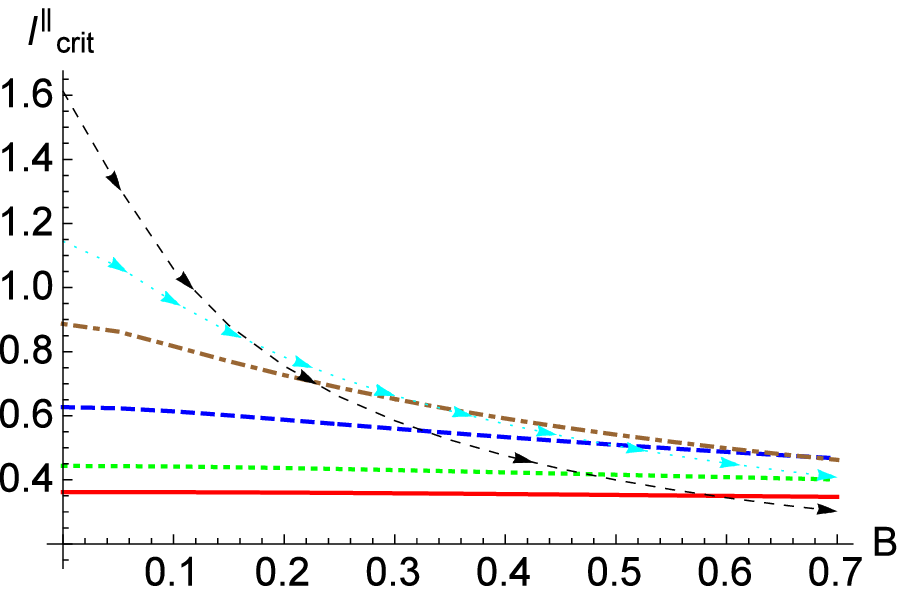}
\caption{\small $\ell_{crit}^{\parallel}$ as a function of $B$ for various $c$. Here $\ell_c=0.2$ and (solid, red), (dot, green), (dash, blue), (dot-dash, brown), (arrow-dot, cyan) and (arrow-dash, black) curves correspond to
$c=3$, $2$, $1$, $0.5$, $0.3$ and $0.15$ respectively. In units GeV.}
\label{lcritvsBlcPt2Zpara}
\end{minipage}
\hspace{0.4cm}
\begin{minipage}[b]{0.5\linewidth}
\centering
\includegraphics[width=2.8in,height=2.3in]{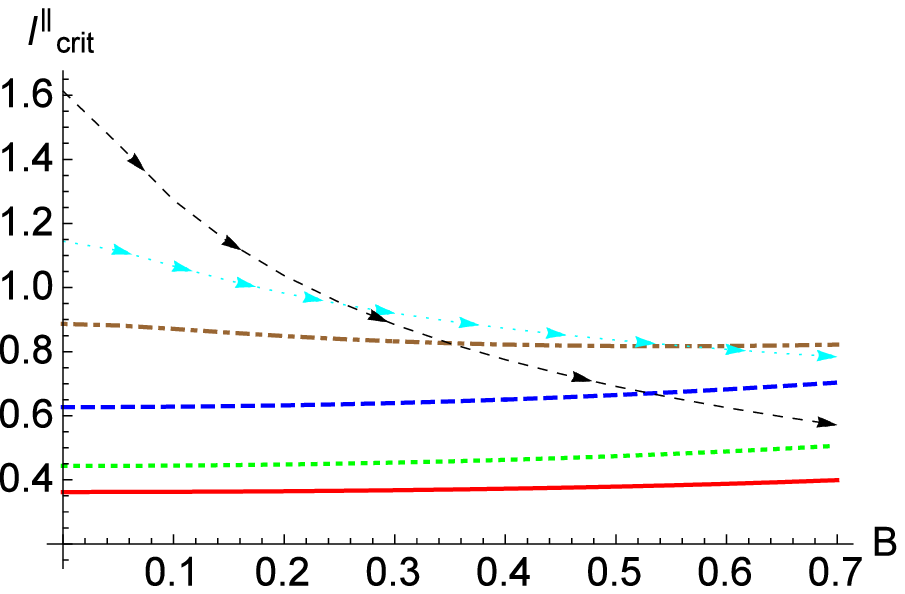}
\caption{\small $\ell_{crit}^{\parallel}$ as a function of $B$ for various $c$. Here $\ell_c=2$ and (solid, red), (dot, green), (dash, blue), (dot-dash, brown), (arrow-dot, cyan) and (arrow-dash, black) curves correspond to
$c=3$, $2$, $1$, $0.5$, $0.3$ and $0.15$ respectively. In units GeV.}
\label{lcritvsBlc2Zpara}
\end{minipage}
\end{figure}

Further, we find that the behaviour of $\Delta S^{\parallel}$ and $\ell_{crit}^{\parallel}$ strongly depends on $\ell_c$. The magnitude and pattern of these quantities with respect to $B$ and $c$ can be different for different $\ell_c$. This can be clearly appreciated from Figures \ref{SvslvsBlcPt2c2Zpara}-\ref{SvslvsBlc2cPt3Zpara}, where $\Delta S^{\parallel}$ as a function of $\ell^{\parallel}$ for $\ell_c=0.2~\text{GeV}^{-1}$ and $\ell_c=2~\text{GeV}^{-1}$ is shown. Here, we have shown the results for two different $c$'s. The dependence of $\ell_{crit}^{\parallel}$ on $B$ and $c$ for these values of $\ell_c$ is shown in Figures \ref{lcritvsBlcPt2Zpara} and \ref{lcritvsBlc2Zpara}. We find that for a fixed $c$ and $B$,  the increase in $\ell_c$ causes $\ell_{crit}^{\parallel}$ to increase, and therefore, $T_{crit}^{\parallel}$ to decrease. This phenomenon is consistent with the confinement/deconfinement phase transition results obtained in \cite{Dudal1511}, where $T_{HP}$ was found to decrease with $\ell_c$. As we will see later on, this result remains valid even for the perpendicular entangling surface. We will say more about this connection at the end of the next subsection.

\subsection{Perpendicular entangling surface}
In this subsection we study the entanglement entropy for the perpendicular entangling surface~\footnote{Again, the constant prefactor $2 L_{y}L_{z}/4 G_{10}$ in the entanglement entropy expression has been dropped from the numerical computation.}. The relevant expressions for the entanglement entropy and strip length are summarized in eqs.~(\ref{SEExparacon})-(\ref{SEExparadis}).\\

\begin{figure}[t!]
\begin{minipage}[b]{0.5\linewidth}
\centering
\includegraphics[width=2.8in,height=2.3in]{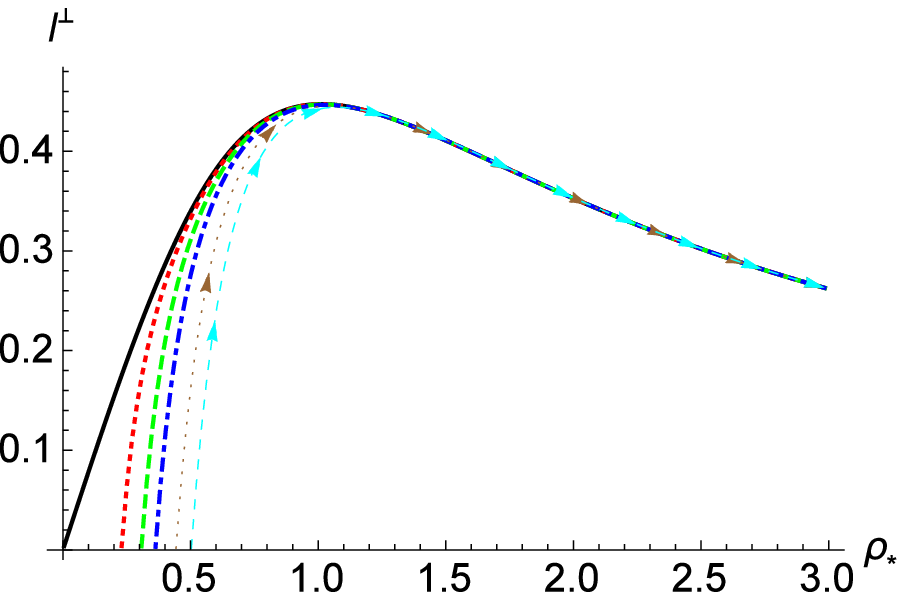}
\caption{ \small $\ell^{\perp}$ as a function of $\rho_*$ for various $B$, with $\ell_c=1$. Here $c=2$ and (solid, black), (dot, red), (dash, green), (dot-dash, blue), (arrow-dot, brown) and (arrow-dash, cyan) curves correspond to $B=0$, $0.1$, $0.2$, $0.3$, $0.5$ and $0.7$ respectively. In units GeV.}
\label{lvsBc2lc1Xpara}
\end{minipage}
\hspace{0.4cm}
\begin{minipage}[b]{0.5\linewidth}
\centering
\includegraphics[width=2.8in,height=2.3in]{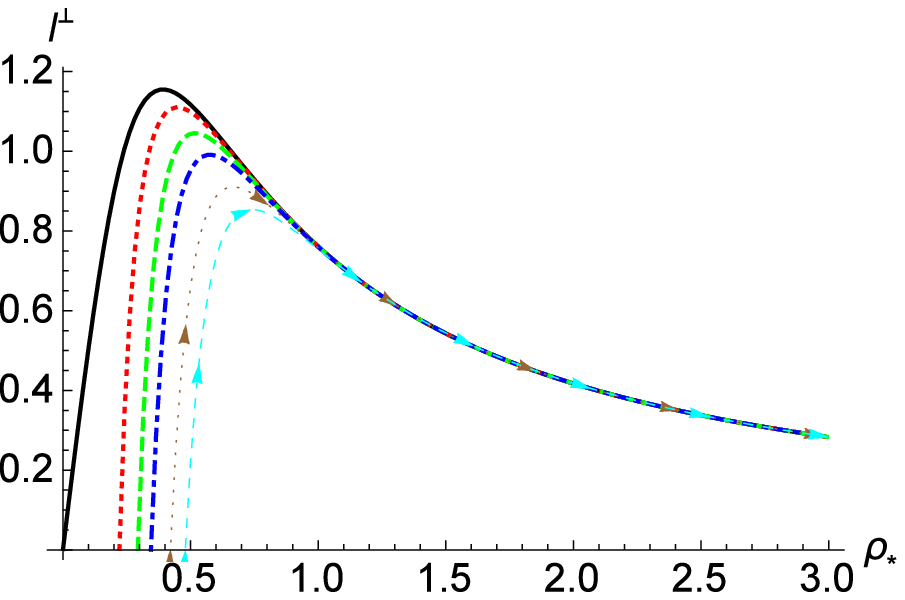}
\caption{\small $\ell^{\perp}$ as a function of $\rho_*$ for various $B$, with $\ell_c=1$. Here  $c=0.3$ and (solid, black), (dot, red), (dash, green), (dot-dash, blue), (arrow-dot, brown) and (arrow-dash, cyan) curves correspond to $B=0$, $0.1$, $0.2$, $0.3$, $0.5$ and $0.7$ respectively. In units GeV.}
\label{lvsBcPt3lc1Xpara}
\end{minipage}
\end{figure}
\begin{figure}[t!]
\begin{minipage}[b]{0.5\linewidth}
\centering
\includegraphics[width=2.8in,height=2.3in]{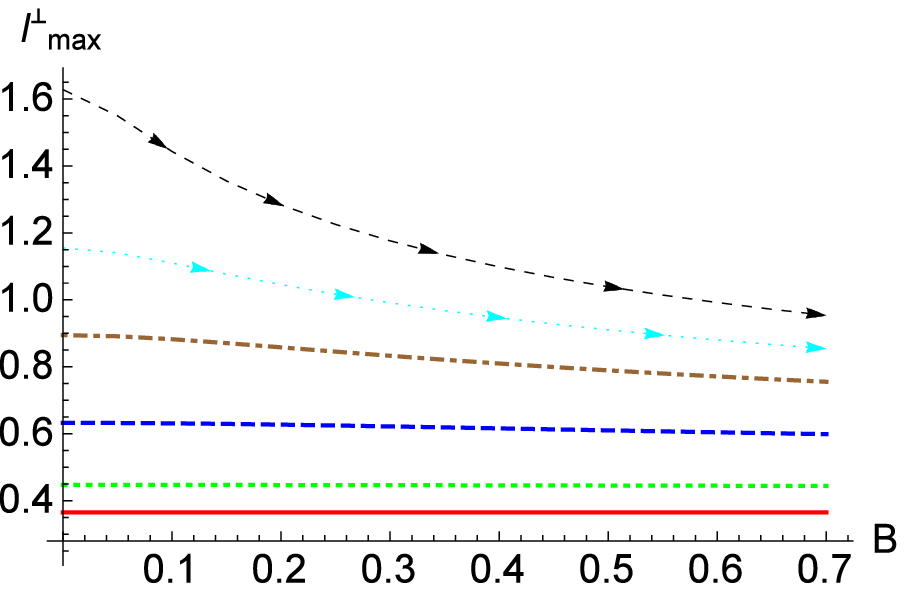}
\caption{ \small $\ell_{max}^{\perp}$ as a function of $B$ for various $c$, with $\ell_c=1$. Here (solid, red), (dot, green), (dash, blue), (dot-dash, brown), (arrow-dot, cyan) and (arrow-dash, black) curves correspond to
$c=3$, $2$, $1$, $0.5$, $0.3$ and $0.15$ respectively. In units GeV.}
\label{lmaxvsBlc1Xpara}
\end{minipage}
\hspace{0.4cm}
\begin{minipage}[b]{0.5\linewidth}
\centering
\includegraphics[width=2.8in,height=2.3in]{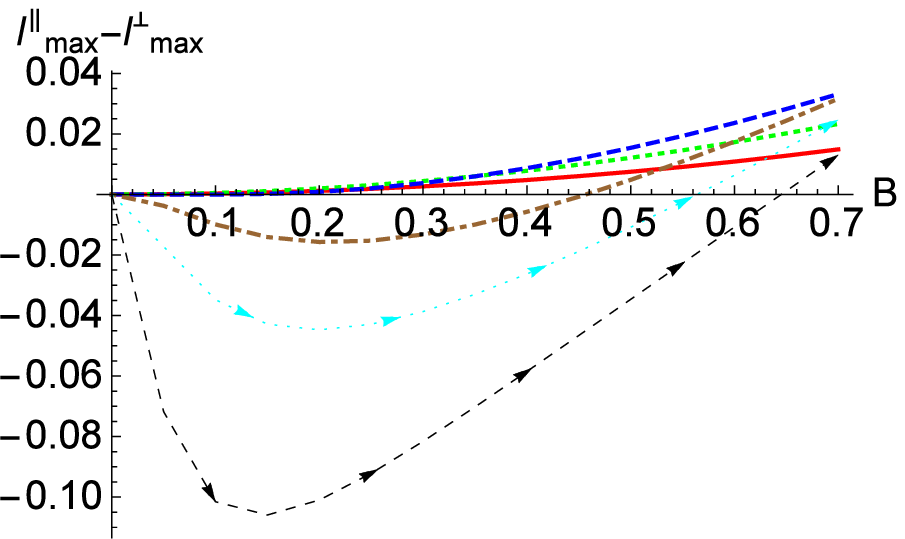}
\caption{\small $\ell_{max}^{\parallel}-l_{max}^{\perp}$ as a function of $B$ for for various $c$, with $\ell_c=1$. Here (solid, red), (dot, green), (dash, blue), (dot-dash, brown), (arrow-dot, cyan) and (arrow-dash, black) curves correspond to
$c=3$, $2$, $1$, $0.5$, $0.3$ and $0.15$ respectively. In units GeV.}
\label{lmaxvsBlc1diff}
\end{minipage}
\end{figure}
Let us first discuss the $\ell_c=1~\text{GeV}^{-1}$ case. The length $\ell^{\perp}$ of the connected surface as function of $\rho_*$ for various values of the magnetic field $B$, for $c=2~\text{GeV}^2$ and $0.3~\text{GeV}^2$, is plotted in Figures \ref{lvsBc2lc1Xpara} and \ref{lvsBcPt3lc1Xpara} respectively. The overall behaviour of $\ell^{\perp}$ is same as was found for $\ell^{\parallel}$. However, there are some differences. Here, $\ell_{max}^{\perp}$ decreases with magnetic field even for $c=2~\text{GeV}^2$ which is opposite to the case of $\ell_{max}^{\parallel}$, which increases with magnetic field. We find that $\ell_{max}^{\perp}$ can be greater or less than $\ell_{max}^{\parallel}$ depending on the values of $B$ and $c$. An overall behaviour is shown in Figures \ref{lmaxvsBlc1Xpara} and \ref{lmaxvsBlc1diff}.\\

\begin{figure}[t!]
\begin{minipage}[b]{0.5\linewidth}
\centering
\includegraphics[width=2.8in,height=2.3in]{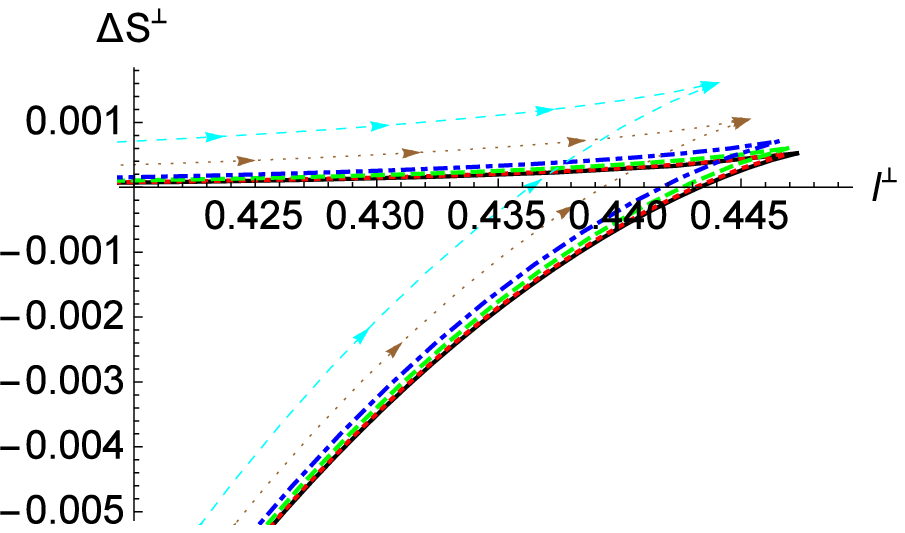}
\caption{ \small $\Delta S^{\perp}$ as a function of $\ell^{\perp}$ for various $B$, with $\ell_c=1$. Here $c=2$ and (solid, black), (dot, red), (dash, green), (dot-dash, blue), (arrow-dot, brown) and (arrow-dash, cyan) curves correspond to $B=0$, $0.1$, $0.2$, $0.3$, $0.5$ and $0.7$ respectively. In units GeV.}
\label{SvslvsBlc1c2Xpara}
\end{minipage}
\hspace{0.4cm}
\begin{minipage}[b]{0.5\linewidth}
\centering
\includegraphics[width=2.8in,height=2.3in]{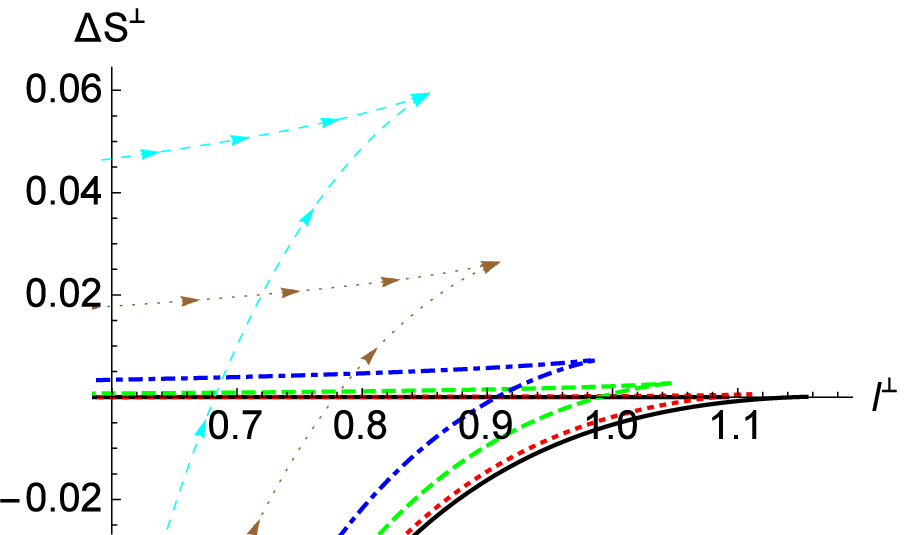}
\caption{\small $\Delta S^{\perp}$ as a function of $\ell^{\perp}$ for various $B$, with $\ell_c=1$. Here $c=0.3$ and (solid, black), (dot, red), (dash, green), (dot-dash, blue), (arrow-dot, brown) and (arrow-dash, cyan) curves correspond to $B=0$, $0.1$, $0.2$, $0.3$, $0.5$ and $0.7$ respectively. In units GeV.}
\label{SvslvsBlc1cPt3Xpara}
\end{minipage}
\end{figure}
\begin{figure}[t!]
\begin{minipage}[b]{0.5\linewidth}
\centering
\includegraphics[width=2.8in,height=2.3in]{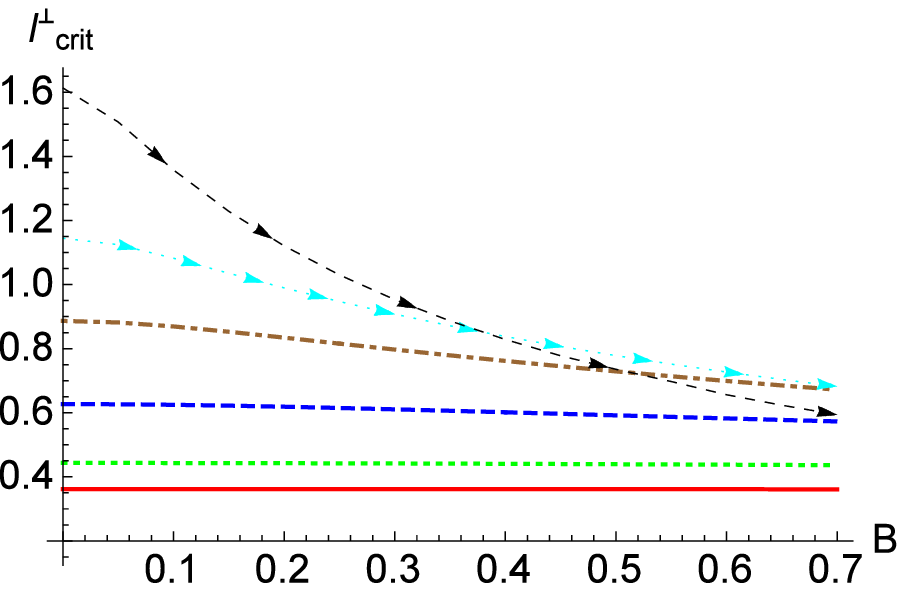}
\caption{\small $\ell_{crit}^{\perp}$ as a function of $B$ for various $c$. Here  $\ell_c=1$ and (solid, red), (dot, green), (dash, blue), (dot-dash, brown), (arrow-dot, cyan) and (arrow-dash, black) curves correspond to
$c=3$, $2$, $1$, $0.5$, $0.3$ and $0.15$ respectively. In units GeV.}
\label{lcritvsBlc1Xpara}
\end{minipage}
\hspace{0.4cm}
\begin{minipage}[b]{0.5\linewidth}
\centering
\includegraphics[width=2.8in,height=2.3in]{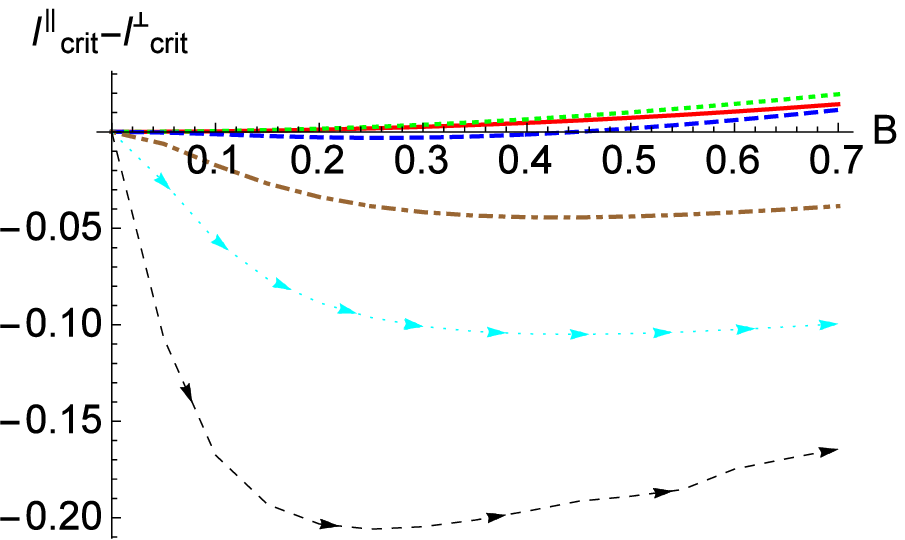}
\caption{\small $\ell_{crit}^{\parallel}-l_{crit}^{\perp}$ as a function of $B$ for various $c$. Here $\ell_c=1$ and (solid, red), (dot, green), (dash, blue), (dot-dash, brown), (arrow-dot, cyan) and (arrow-dash, black) curves correspond to
$c=3$, $2$, $1$, $0.5$, $0.3$ and $0.15$ respectively. In units GeV.}
\label{lcritvsBlc1diff}
\end{minipage}
\end{figure}
\begin{figure}[t!]
\begin{minipage}[b]{0.5\linewidth}
\centering
\includegraphics[width=2.8in,height=2.3in]{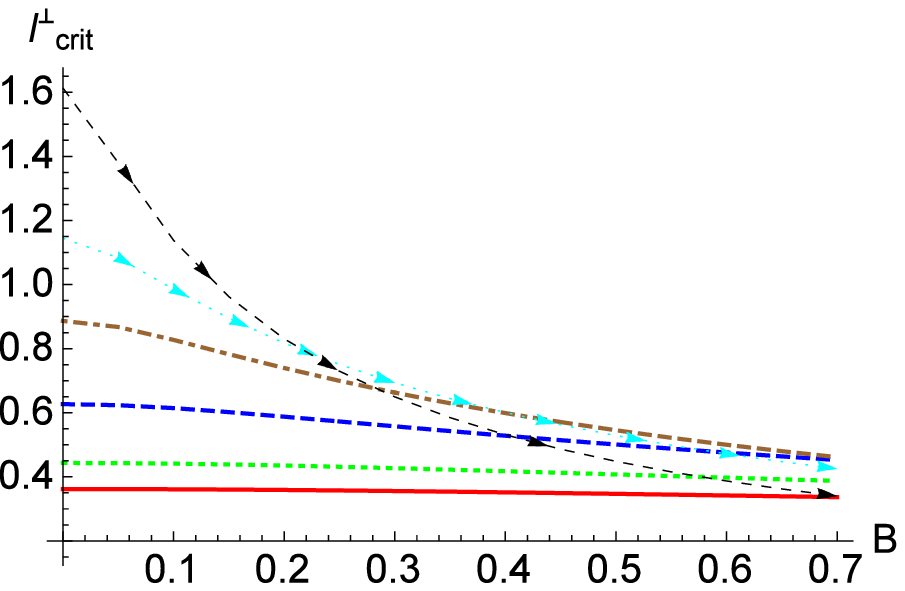}
\caption{\small $\ell_{crit}^{\perp}$ as a function of $B$ for various $c$. Here  $\ell_c=0.2$ and (solid, red), (dot, green), (dash, blue), (dot-dash, brown), (arrow-dot, cyan) and (arrow-dash, black) curves correspond to
$c=3$, $2$, $1$, $0.5$, $0.3$ and $0.15$ respectively. In units GeV.}
\label{lcritvsBlcPt2Xpara}
\end{minipage}
\hspace{0.4cm}
\begin{minipage}[b]{0.5\linewidth}
\centering
\includegraphics[width=2.8in,height=2.3in]{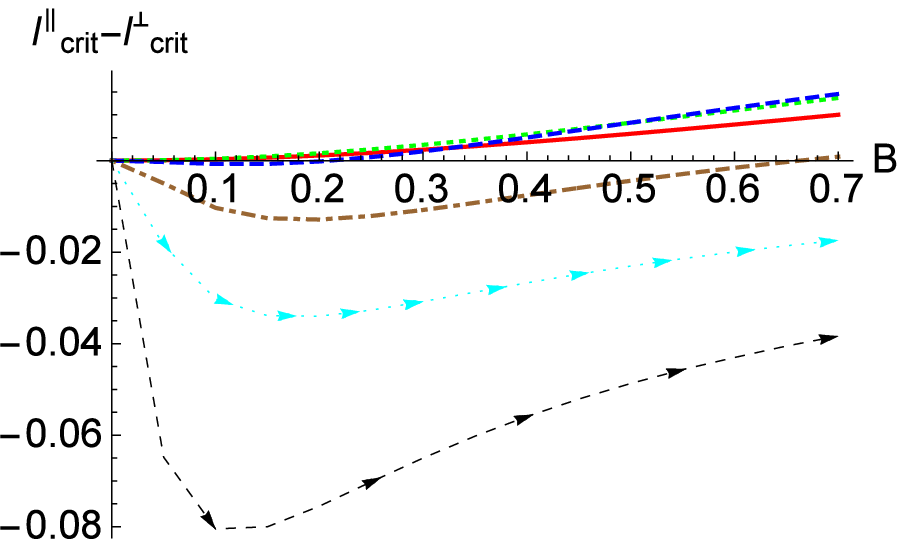}
\caption{\small $\ell_{crit}^{\parallel}-l_{crit}^{\perp}$ as a function of $B$ for for various $c$. Here $\ell_c=0.2$ and (solid, red), (dot, green), (dash, blue), (dot-dash, brown), (arrow-dot, cyan) and (arrow-dash, black) curves correspond to
$c=3$, $2$, $1$, $0.5$, $0.3$ and $0.15$ respectively. In units GeV.}
\label{lcritvsBlcPt2diff}
\end{minipage}
\end{figure}
\begin{figure}[t!]
\begin{minipage}[b]{0.5\linewidth}
\centering
\includegraphics[width=2.8in,height=2.3in]{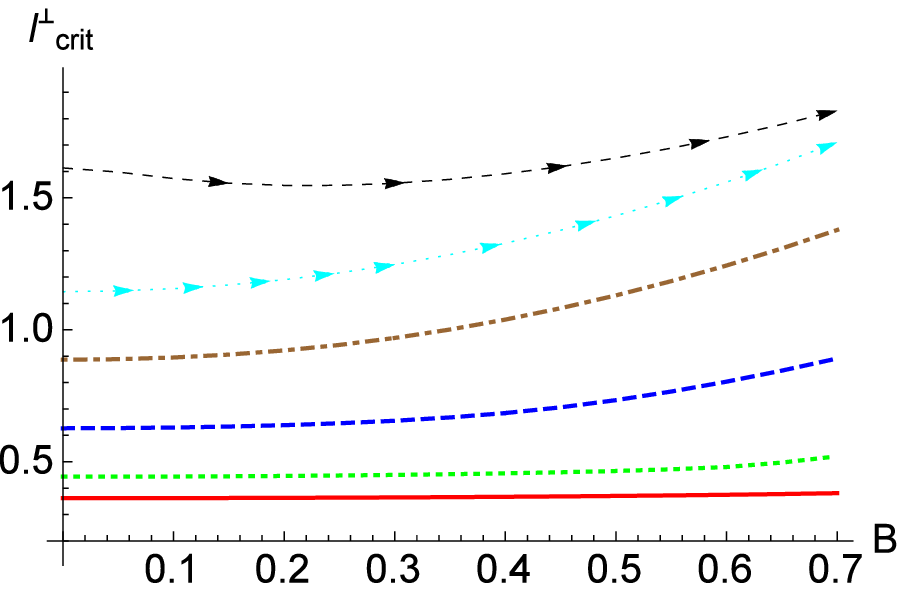}
\caption{\small $\ell_{crit}^{\perp}$ as a function of $B$ for various $c$. Here  $\ell_c=2$ and (solid, red), (dot, green), (dash, blue), (dot-dash, brown), (arrow-dot, cyan) and (arrow-dash, black) curves correspond to
$c=3$, $2$, $1$, $0.5$, $0.3$ and $0.15$ respectively. In units GeV.}
\label{lcritvsBlc2Xpara}
\end{minipage}
\hspace{0.4cm}
\begin{minipage}[b]{0.5\linewidth}
\centering
\includegraphics[width=2.8in,height=2.3in]{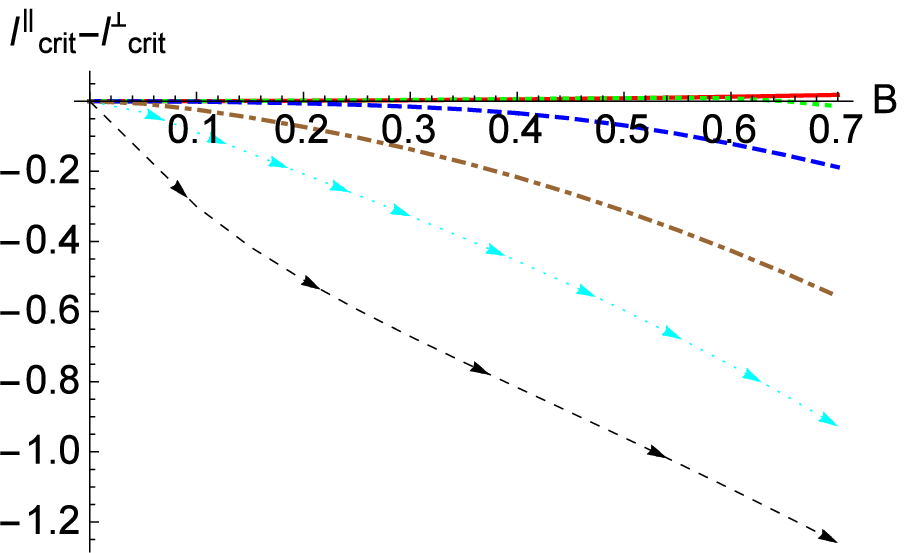}
\caption{\small $\ell_{crit}^{\parallel}-l_{crit}^{\perp}$ as a function of $B$ for for various $c$. Here $\ell_c=2$ and (solid, red), (dot, green), (dash, blue), (dot-dash, brown), (arrow-dot, cyan) and (arrow-dash, black) curves correspond to
$c=3$, $2$, $1$, $0.5$, $0.3$ and $0.15$ respectively. In units GeV.}
\label{lcritvsBlc2diff}
\end{minipage}
\end{figure}

The difference between connected and disconnected entanglement entropies is shown in Figures \ref{SvslvsBlc1c2Xpara} and \ref{SvslvsBlc1cPt3Xpara}. Even with the perpendicular entangling surface, we find that $\Delta S^{\perp}$ can be greater or less than zero and that a phase transition from connected to disconnected surface occur as we increase the strip length $\ell^{\perp}$. The critical length at which this phase transition occur is now defined as $\ell_{crit}^{\perp}$.  An important point to observe is that $\ell_{crit}^{\perp}$ for $c=2~\text{GeV}^2$ decreases with the magnetic field in contrast with the case of $\ell_{crit}^{\parallel}$, which increases with the magnetic field. The dependence of $\ell_{crit}^{\perp}$ on $B$ and $c$ is shown in Figure \ref{lcritvsBlc1Xpara}, which is quite distinct compared to the behaviour of $\ell_{crit}^{\parallel}$ (shown in Figure \ref{lcritvsBlc1Zpara}), especially for smaller values of $c$ and larger values of $B$. The difference between $\ell_{crit}^{\parallel}$ and $\ell_{crit}^{\perp}$ is shown in Figure \ref{lcritvsBlc1diff}. We see that $\ell_{crit}^{\parallel}>\ell_{crit}^{\perp}$ for larger values of $c$, however as we decrease the dilaton factor $c$ to near the QCD value ($c=0.3~\text{GeV}^2$), we find $\ell_{crit}^{\parallel}<\ell_{crit}^{\perp}$. This suggests that $T_{crit}^{\parallel}>T_{crit}^{\perp}$ in the boundary QCD theory. For $B=0$, we find the expected result $\ell_{crit}^{\parallel}=\ell_{crit}^{\perp}$.\\
\begin{figure}[t!]
\begin{minipage}[b]{0.5\linewidth}
\centering
\includegraphics[width=2.8in,height=2.3in]{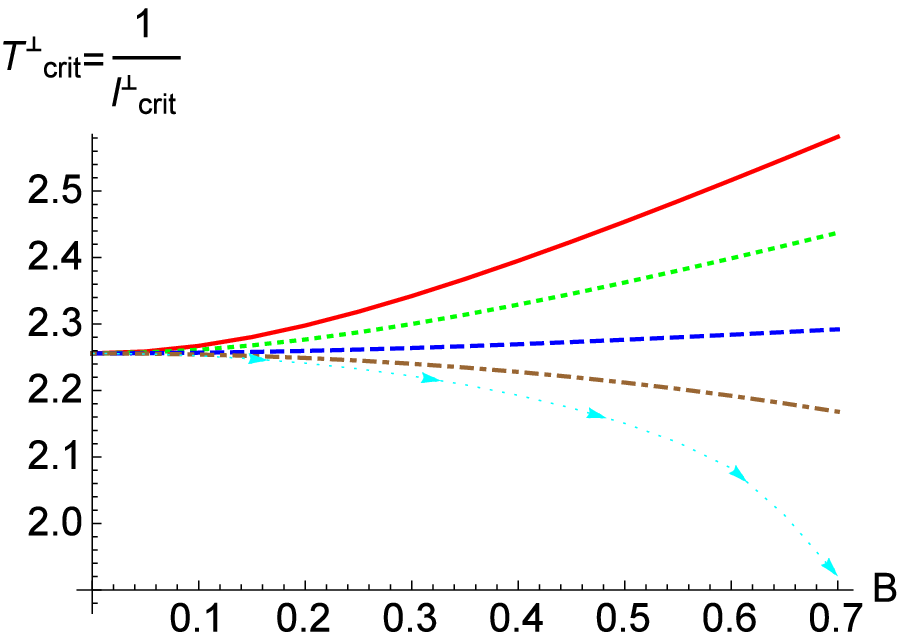}
\caption{\small $T_{crit}^{\perp}$ as a function of $B$ for various $\ell_c$. Here  $c=2$ and (solid, red), (dot, green), (dash, blue), (dot-dash, brown) and (arrow-dot, cyan) curves correspond to $\ell_c=0.2$, $0.5$, $1$, $1.5$ and $2$ respectively. In units GeV.}
\label{lcritvsBvslcc2Xpara}
\end{minipage}
\hspace{0.4cm}
\begin{minipage}[b]{0.5\linewidth}
\centering
\includegraphics[width=2.8in,height=2.3in]{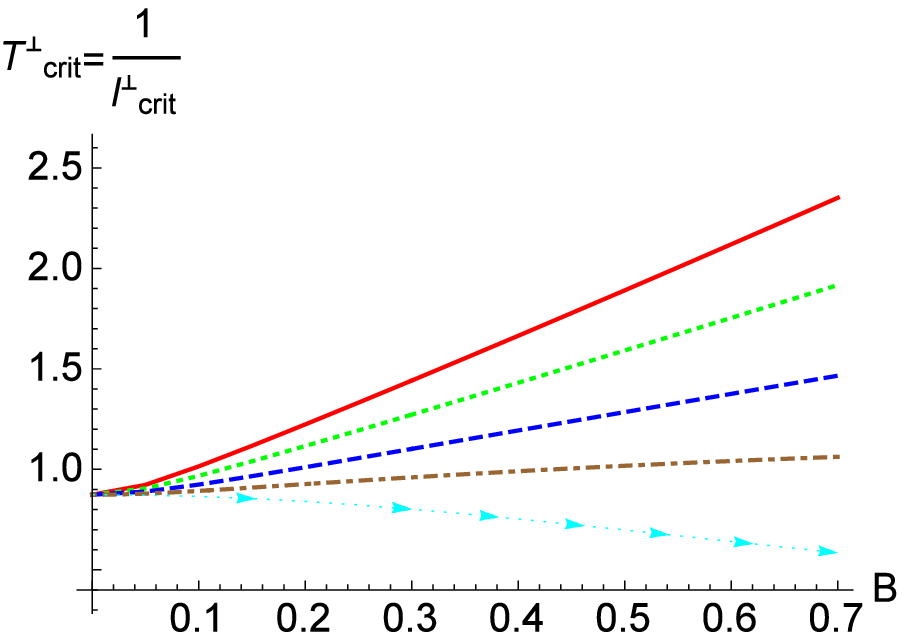}
\caption{\small $T_{crit}^{\perp}$ as a function of $B$ for various $\ell_c$. Here  $c=0.3$ and (solid, red), (dot, green), (dash, blue), (dot-dash, brown) and (arrow-dot, cyan) curves correspond to $\ell_c=0.2$, $0.5$, $1$, $1.5$ and $2$ respectively. In units GeV.}
\label{lcritvsBvslccPt3Xpara}
\end{minipage}
\end{figure}

Of course the above analysis is also sensitive to the length scale $\ell_c$. For two different values of $\ell_c$, the results for $\ell_{crit}^{\perp}$ are shown in Figures \ref{lcritvsBlcPt2Xpara}-\ref{lcritvsBlc2diff}. One can clearly notice the changes in the pattern of $\ell_{crit}^{\perp}$ as we vary $\ell_c$. In particular, we can notice that $\ell_{crit}^{\perp}$ shows monotonic behaviour with respect to $c$ even for higher magnetic field as we make $\ell_c$ larger and larger (Figure \ref{lcritvsBlc2Xpara}). As in the case of $\ell_{crit}^{\parallel}$, here too we find that for a fixed $c$ and $B$,  the increase in $\ell_c$ causes $\ell_{crit}^{\perp}$ to increase. Similarly, $\ell_{crit}^{\perp}$ can be greater or less than $\ell_{crit}^{\parallel}$ depending on the values of $B$ and $c$. In particular, for larger $\ell_c$, $\ell_{crit}^{\perp}$ generally dominates $\ell_{crit}^{\parallel}$.\\

It is interesting to connect our results of entanglement entropy with the free energy results of \cite{Dudal1511} and analyze the similarities and differences between them. For this purpose, a few points are in order:
\begin{itemize}
\item In \cite{Dudal1511}, it was shown that the critical temperature $T_{HP}$ of the Hawking-Page (i.e.~the dual of the confinement/deconfinement phase transition) decreases with $\ell_c$. Considering that the strip length of the entangling surface plays the role of inverse temperature, the corresponding critical temperature in the entanglement entropy also shows the same feature. This is shown in Figures \ref{lcritvsBvslcc2Xpara} and \ref{lcritvsBvslccPt3Xpara}, where one clearly see that for any fixed $B$ and $c$, $T_{crit}^{\perp}$ decreases as the value of $\ell_c$ increases. As briefly mentioned in the previous subsection, the same result is true for $T_{crit}^{\parallel}$ as well.

\item There are several differences between $T_{HP}$ and $T_{crit}^{\perp}$ as well, especially as a function of $B$ and $c$. $T_{HP}$ can decrease or increase with magnetic field depending on the values of $c$ and $\ell_c$. In \cite{Dudal1511}, it was found that for $c=0.3~\text{GeV}^2$ (which in the notation of \cite{Dudal1511} corresponds to $c=0.15~\text{GeV}^2$), $T_{HP}$ increases with magnetic field for smaller $\ell_c$, say $\ell_c<1~\text{GeV}^{-1}$, and decreases for larger $\ell_c$. However, the same feature does not occur with $T_{crit}^{\perp}$. This can be seen from Figure \ref{lcritvsBvslccPt3Xpara}, where we find that $T_{crit}^{\perp}$ decreases with magnetic field only for $\ell_c\gtrapprox 1.7~\text{GeV}^{-1}$. Similar results hold for other values of $c$ as well. Therefore, it seems that the relation between $T_{crit}^{\perp}$ and $T_{HP}$ is rather complicated and a straightforward one to one comparison between them is bit subtle.

\item Analogous differences exist for $T_{crit}^{\parallel}$ as well. Again, we do not find a straightforward one to one relation between $T_{HP}$ and $T_{crit}^{\parallel}$.
\end{itemize}

We end this subsection by presenting a qualitative summery of our results showing how $\ell_{crit}^{\parallel}$ and $\ell_{crit}^{\perp}$ change with magnetic field for different values of the $c$ and $\ell_c$. This is succinctly shown in Table~\ref{tablelcrit}.

\begin{table}[ht]
\begin{center}
 \begin{tabular}{|c|c|c|c|c|c|}
\multicolumn{6}{c}{}\\
\hline
\backslashbox{$c$}{$\ell_c$} & $0.2$ & $0.5$ & $1$ & $1.5$ & $2$ \\
\hline
\vspace{0.1mm}
 & $\parallel$ \hspace{1cm} $\perp$ & $\parallel$ \hspace{1cm}  $\perp$ &$\parallel$ \hspace{1cm}  $\perp$ & $\parallel$ \hspace{1cm}  $\perp$ & $\parallel$ \hspace{1cm}  $\perp$
\\
\hline
\vspace{0.1mm}
$3$ &$\downarrow$ \hspace{1cm} $\downarrow$ & $\downarrow$ \hspace{1cm} $\downarrow$ & $\uparrow$ \hspace{1cm} $\downarrow$ & $\uparrow$ \hspace{1cm} $\uparrow$ & $\uparrow$ \hspace{1cm}  $\uparrow$
\\
\hline
\vspace{0.1mm}
$2$ &$\downarrow$ \hspace{1cm} $\downarrow$ & $\downarrow$ \hspace{1cm} $\downarrow$ & $\uparrow$ \hspace{1cm} $\downarrow$ & $\uparrow$ \hspace{1cm} $\uparrow$ & $\uparrow$ \hspace{1cm}  $\uparrow$
\\
\hline
\vspace{0.1mm}
$1$ &$\downarrow$ \hspace{1cm} $\downarrow$ & $\downarrow$ \hspace{1cm} $\downarrow$ & $\downarrow$ \hspace{1cm} $\downarrow$ & \small{NM} \hspace{0.6cm} $\uparrow$ & $\uparrow$ \hspace{1cm}  $\uparrow$ \\
\hline
\vspace{0.1mm}
$0.5$ &$\downarrow$ \hspace{1cm} $\downarrow$ & $\downarrow$ \hspace{1cm} $\downarrow$ & $\downarrow$ \hspace{1cm} $\downarrow$ & $\downarrow$ \hspace{1cm} $\downarrow$ & \small{NM} \hspace{0.6cm}  $\uparrow$ \\
\hline
\vspace{0.1mm}
$0.3$ &$\downarrow$ \hspace{1cm} $\downarrow$ & $\downarrow$ \hspace{1cm} $\downarrow$ & $\downarrow$ \hspace{1cm} $\downarrow$ & $\downarrow$ \hspace{1cm} $\downarrow$ & $\downarrow$ \hspace{1cm}  $\uparrow$ \\
\hline
\vspace{0.1mm}
$0.15$ & $\downarrow$ \hspace{1cm} $\downarrow$ & $\downarrow$ \hspace{1cm} $\downarrow$ & $\downarrow$ \hspace{1cm} $\downarrow$ & $\downarrow$ \hspace{1cm} $\downarrow$ & $\downarrow$ \hspace{0.6cm}  \small{NM} \\
\hline
\end{tabular}
\caption{A summary of the dependence of the entanglement entropy on the background magnetic field for various values of $c$ and $\ell_c$. Symbols $\parallel$ or $\perp$ indicate results for parallel or perpendicular entangling surfaces. Arrows $\uparrow$ or $\downarrow$ indicate whether $\ell_{crit}^{\parallel}$ or $\ell_{crit}^{\perp}$ increases or decreases as we increase the magnetic field. Here NM stands for non-monotonic and indicates that the concerned quantity first decreases and then increases with the magnetic field. The parameters $c$ and $\ell_c$ at which the holographic model in eq.~(\ref{action}) is best suitable to describe genuine QCD correspond to $c\simeq 0.3~\text{GeV}^{2}$ and $l_c\gtrapprox 1~\text{GeV}^{-1}$.}
\label{tablelcrit}
\end{center}
\end{table}

\subsection{The entropic $\cal C$-function}
\begin{figure}[t!]
\begin{minipage}[b]{0.5\linewidth}
\centering
\includegraphics[width=2.8in,height=2.3in]{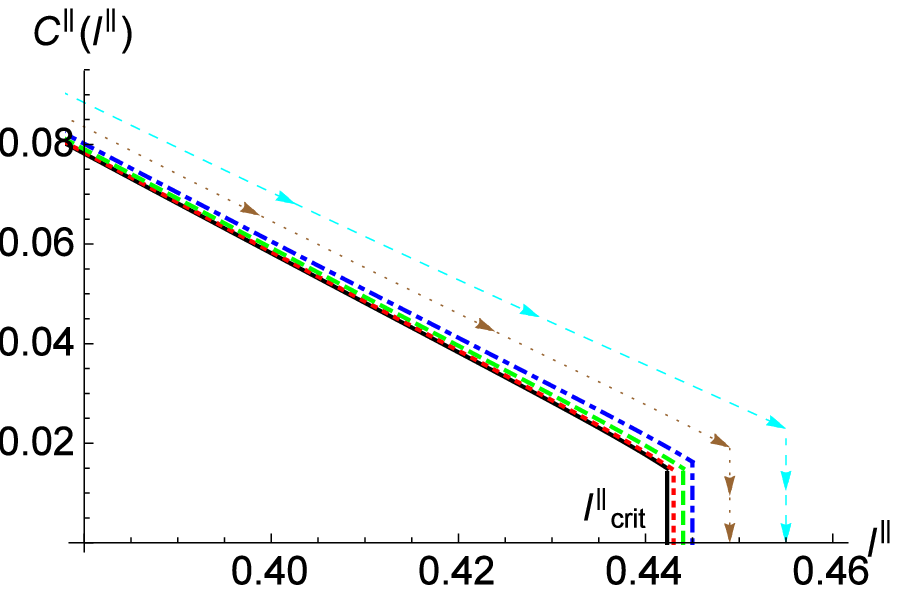}
\caption{\small Entropic $\cal C$-function for a parallel entangling surface as a function of length $\ell^\parallel$. Here  $c=2$ and (solid, black), (dot, red), (dash, green), (dot-dash, blue), (arrow-dot, brown) and (arrow-dash, cyan) curves correspond to $B=0$, $0.1$, $0.2$, $0.3$, $0.5$ and $0.7$ respectively. In units GeV.}
\label{CfuncvsBlc1c2Zpara}
\end{minipage}
\hspace{0.4cm}
\begin{minipage}[b]{0.5\linewidth}
\centering
\includegraphics[width=2.8in,height=2.3in]{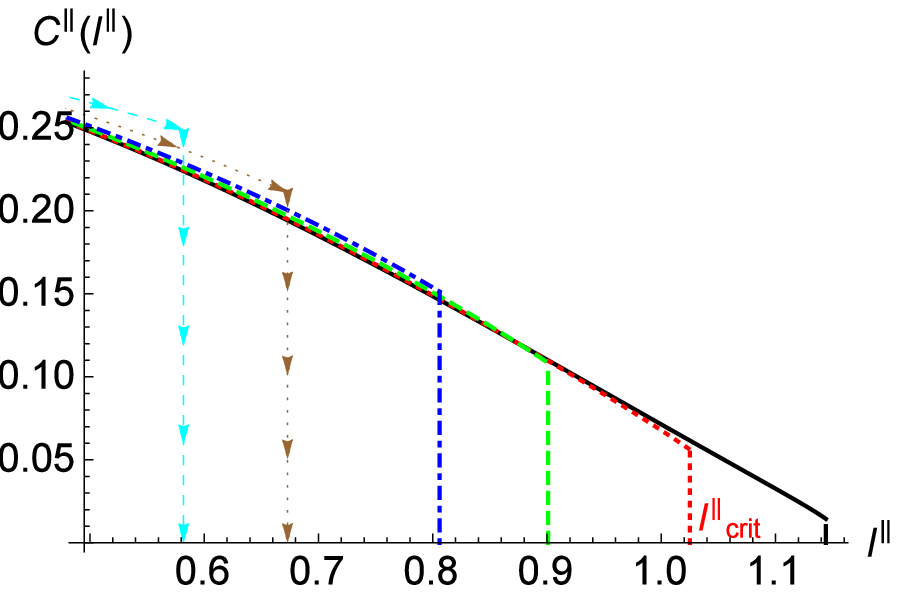}
\caption{\small Entropic $\cal C$-function for a parallel entangling surface as a function of length $\ell^\parallel$. Here  $c=0.3$ and (solid, black), (dot, red), (dash, green), (dot-dash, blue), (arrow-dot, brown) and (arrow-dash, cyan) curves correspond to $B=0$, $0.1$, $0.2$, $0.3$, $0.5$ and $0.7$ respectively. In units GeV.}
\label{CfuncvsBlc1cPt3Zpara}
\end{minipage}
\end{figure}
\begin{figure}[t!]
\begin{minipage}[b]{0.5\linewidth}
\centering
\includegraphics[width=2.8in,height=2.3in]{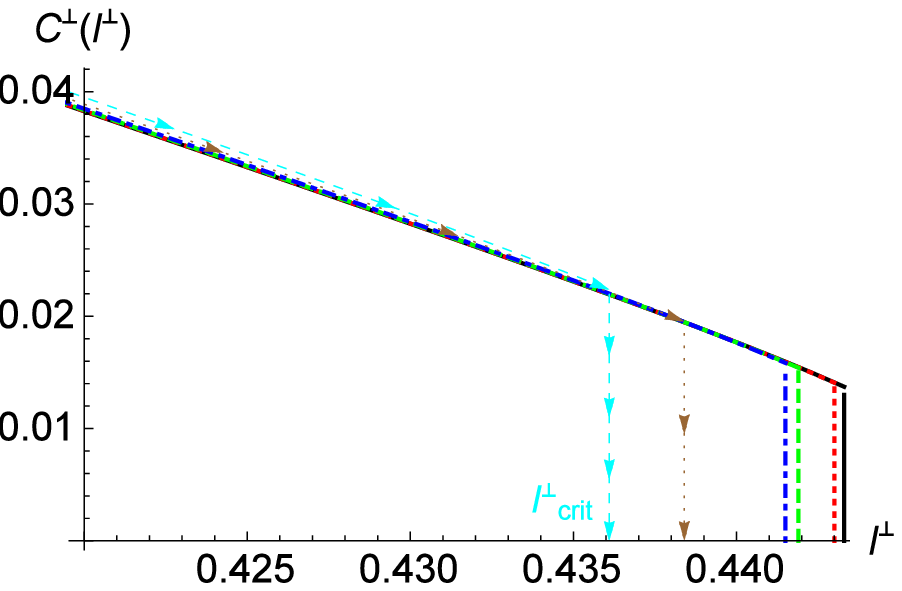}
\caption{\small Entropic $\cal C$-function for a perpendicular entangling surface as a function of length $\ell^\perp$. Here  $c=2$ and (solid, black), (dot, red), (dash, green), (dot-dash, blue), (arrow-dot, brown) and (arrow-dash, cyan) curves correspond to $B=0$, $0.1$, $0.2$, $0.3$, $0.5$ and $0.7$ respectively. In units GeV.}
\label{CfuncvsBlc1c2Xpara}
\end{minipage}
\hspace{0.4cm}
\begin{minipage}[b]{0.5\linewidth}
\centering
\includegraphics[width=2.8in,height=2.3in]{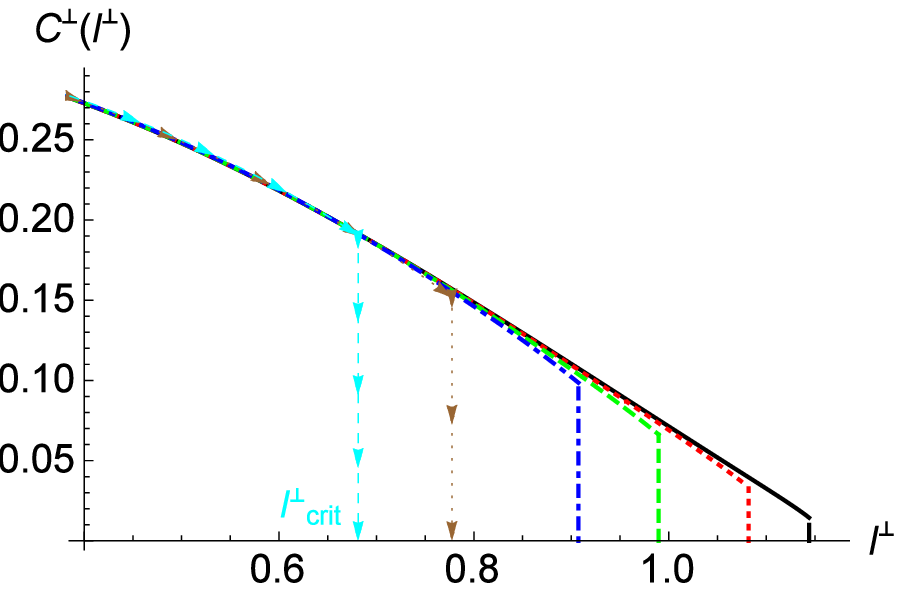}
\caption{\small Entropic $\cal C$-function for a perpendicular entangling surface as a function of length $\ell^\perp$. Here  $c=0.3$ and (solid, black), (dot, red), (dash, green), (dot-dash, blue), (arrow-dot, brown) and (arrow-dash, cyan) curves correspond to $B=0$, $0.1$, $0.2$, $0.3$, $0.5$ and $0.7$ respectively. In units GeV.}
\label{CfuncvsBlc1cPt3Xpara}
\end{minipage}
\end{figure}
In this subsection, we briefly discuss our results for the entropic $\mathcal{C}$-function, which on general grounds is defined as \cite{Nishioka0611,Itou:2015cyu}
\begin{eqnarray}
{\cal C}(\ell)=\frac{\ell^3}{\text{Area}(A)}\frac{\partial S}{\partial \ell}\,,
\label{Cfunc}
\end{eqnarray}
where $\text{Area}(A)$ is the area of the subsystem $A$. By construction, eq.~\eqref{Cfunc} is finite. In our case, there can be two entropic $\cal C$-functions depending on whether the entangling strip is parallel or perpendicular to the magnetic field. The results for the parallel case are shown in Figures \ref{CfuncvsBlc1c2Zpara} and \ref{CfuncvsBlc1cPt3Zpara}, where we have chosen $c=2~\text{GeV}^2$ and $c=0.3~\text{GeV}^2$ respectively. We see that the magnitude of $\mathcal{C}^\parallel$ decreases monotonically as we increase the length of the strip, i.e.~from UV to IR. Since, $\ell^\parallel$ is inversely related to the energy scale of the theory and that $\cal C$ measures the degrees of freedom at that energy scale \cite{Nishioka0611}, this result is consistent with the expected behaviour of $\cal C$ that it decreases under the RG-flow in a confining theory. The $\cal C$-function sharply drops to zero at $\ell_{crit}^\parallel$, indicative of a first order transition, and it continues to vanish for higher $\ell^\parallel$. This is precisely due to the reason that for $\ell^\parallel > \ell_{crit}^\parallel$, the entanglement entropy of the connected surface dominates that of the disconnected surface and that the entanglement entropy of the disconnected surface is independent of $\ell^\parallel$.\\

The entropic $\cal C$-function for a perpendicular entangling surface shows a similar behaviour and is shown in Figures \ref{CfuncvsBlc1c2Xpara} and \ref{CfuncvsBlc1cPt3Xpara}.

\section{Outlook}
We have set a next modest step in further unraveling the ``entanglement'' between confinement and entanglement entropy, this to further understand the intricacies of confinement in QCD when a magnetic field is introduced as a classic background. This is of phenomenological relevance to quark-gluon plasma physics, as advocated in many quoted papers, for example to understand how (confined) heavy quark bound states will react if the temperature is sufficiently high and a strong magnetic field is present. The latter is presumably generated due to the (non-central) heavy ion collision leading to the plasma phase.\\

We have given first evidence that the entanglement entropy feels the magnetic field $\vec{B}=B\vec{e}_z$ in two ways: the critical lengths of the entangling strip surfaces become not only $B$-dependent, indicating a phase transition that is $B$-dependent, but the critical length depends also on the either parallel or perpendicular orientation of the surface with respect to $\vec{B}$. For definiteness, we did not consider a general angle $\theta$ between surface and $\vec{B}$ to avoid having to deal with yet another parameter.\\

However, several questions remain. The most pertinent one would be to clarify the link between the anisotropy in the confining behaviour, as signalled by the entanglement entropy structure, and that signalled by the string tensions extracted from a Wilson loop \cite{Bonati:2014ksa,Bonati:2016kxj} in a magnetic background. It is expected, at least from a holographic viewpoint, that there is an intimate connection between the area law of the Wilson loop, viz.~confinement, and the behaviour of the entanglement entropy \cite{Kola1403}. Though, to put this on a firmer footing, we believe we need to first ensure that the necessary dilaton factor, with or without magnetic field, is coupled to the theory in a self-consistent way, that is, by solving the bulk Einstein equations of motion, while simultaneously ensuring the area law for the holographic representation of the Wilson loop. Such approach can possibly also help to get a better handle over the length parameter $\ell_c$ that enters the metric at finite $B$ and which connection to real QCD is still a bit mystified. Once this is done, we can move forward to study the entanglement entropy in such improved setting. It will also allow to identify, even for zero magnetic field, the r\^{o}le, if any, of entanglement entropy in the so-called entropic destruction picture of the dissociation of a heavy quark bound state \cite{Kharzeev:2014pha,Hashimoto:2014fha}. We plan to come back to these issues in future work.

\section*{Acknowledgments}
We thank T.~Mertens and A.~Blommaert for useful discussions. S.~Mahapatra is supported by a PDM grant of KU Leuven.

\end{document}